\pdfoutput=1
\documentclass[twocolumn,english,aps,longbibliography,superscriptaddress,prb]{revtex4-2}

\usepackage{babel}
\usepackage{amsmath}
\usepackage{amssymb}
\usepackage{graphicx}
\usepackage{xcolor}
\usepackage{tikz}
\definecolor{airforceblue}{rgb}{0.36, 0.54, 0.66}
\definecolor{colorPT}{rgb}{0.93, 0.69, 0.13}
\definecolor{darkcoral}{rgb}{0.85, 0.33, 0.10}
\definecolor{colorKT}{rgb}{0.00, 0.45, 0.74}

\begin{document}

\title{Commensurate-incommensurate transition in the chiral Ashkin-Teller model}

\author{Samuel Nyckees}
\affiliation{Institute of Physics, Ecole Polytechnique F\'ed\'erale de Lausanne (EPFL), CH-1015 Lausanne, Switzerland}
\author{Fr\'ed\'eric Mila}
\affiliation{Institute of Physics, Ecole Polytechnique F\'ed\'erale de Lausanne (EPFL), CH-1015 Lausanne, Switzerland}

\date{\today}
\begin{abstract}

We investigate the classical chiral Ashkin-Teller model on a square lattice with the corner transfer matrix renormalisation group (CTMRG) algorithm. We 
show that the melting of the period-4 phase in the presence of a chiral perturbation takes different forms depending on the coefficient of the four-spin term in the Ashkin-Teller model.
Close to the clock limit of two decoupled Ising models, the system undergoes a two-step commensurate-incommensurate transition as soon as the chirality is introduced, with an intermediate critical floating phase bounded by a Kosterlitz-Thouless transition at high temperature and a Pokrovsky-Talapov transition at low temperature. By contrast, close to the four-states Potts model, we argue for the existence of a unique commensurate-incommensurate transition that belongs to the chiral universality class, and for the presence of a Lifshitz point where the ordered, disordered and floating phases meet. Finally, we map the whole phase diagram, which turns out to be in qualitative agreement with the 40 year old prediction by Huse and Fisher.

\end{abstract}

\maketitle

%%%%%%%%%%%%%%%%%%%%%%%%%%%%%%%%%%%%% INTRODUCTION %%%%%%%%%%%%%%%%%%%%%%%%%%%%%%%%%%%%

\section{Introduction}

Recent experiments on Rydberg atoms\cite{lukin2017,lukin2019} have brought back the problem of commensurate-incommensurate (C-IC) transitions, and of the nature of the melting of an ordered period-$p$ phase into an incommensurate one. This transition was originally discussed in two dimensional classical systems in the context of the melting of an adsorbed layer on a substrate lattice, and simultaneously introduced by Huse\cite{Huse1981} and Ostlund\cite{Ostlund} as a series of chiral models exhibiting $p-1$ types of domain walls, such that domain walls between ordered domains $A\mid B$ and $B\mid A$ have different energy, introducing a chiral into the problem. %Therefore, providing toy models for the melting of an adsorbed surface on a substrate lattice.
For $p=2$, the system is commensurate even in the disordered phase, and there is no commensurate-incommensurate transition. The transition, if continuous, generically belongs to the Ising universality class. 
For $p\geq 5$ the transition is always a two-step one, with a critical phase in between. In that phase, the configurations are made of permutations of stripes of the $p$ different orders with domain walls at an average distance of $l\propto1/q$, where $q$ is the incommensurate wave-vector of the correlations. As the temperature approaches the ordered phase, the domain walls repel each other and the wave-vector converges to a commensurate value with an exponent $\bar{\beta}$ ($q - q_0 \sim t^{\bar{\beta}}$, where $t=(T-T_c)/T_c$ is the reduced temperature). It is commonly agreed that the transition is in the Pokrovsky-Talapov (PT)\cite{Pokrovsky_Talapov} universality class, characterised by critical exponents : $\nu_x= 1/2, \nu_y=1,\bar{\beta}=1/2$, where $\nu_{x/y}$ describes the divergence of the correlation length in the directions $x$ and $y$ respectively. The high temperature transition between the critical and the disordered phases is a Kosterlitz-Thouless\cite{Kosterlitz_Thouless_1973} one driven by the unbinding of the dislocations\cite{JoseKadanoff77} which become relevant at $\eta=1/4$, where $\eta$ is the exponent of the decay of the spin-spin correlation function. 

For the $p=3,4$ cases, the picture might be different. It was first argued by Huse and Fisher \cite{HuseFisher} that a floating critical phase bounded by two transitions might not appear right away when introducing the chirality. Instead the commensurate-incommensurate transition could be unique and would belong to a new chiral universality class characterised by the product of the correlation length $\xi$ and the wave-vector $q-q_0$ converging to a constant at criticality ($\overline{\beta}=\nu$), and by a dynamical (or anisotropy) exponent $z\equiv \nu_y/\nu_x$ different from 1 ($\nu_y\neq\nu_x$). 

Since its first introduction, the existence of the chiral universality class has generated numerous studies, especially for the $p=3$ case incarnated by the three-state chiral Potts model. Both the classical \cite{HuseFisher1984,HuseFisher,schulz1980,schulz,cardy,AUYANG199678,yeomans1985,sato_sasaki,Selke1982,Duxbury,sato,houlrik1986,auyang1996,auyang1987,baxter1988} and quantum \cite{howes1983,chepiga_mila_PRL,samajdar,Everts_1989,hughes,HOWES1983169,sachdev_dual,CENTEN1982585} cases have been studied extensively. Further experimental work was also performed\cite{bartelt,abernathy,SelkeExperiment}. 

By contrast to the $p=3$ case, the $p=4$ literature is less rich. We note however that recently an investigation of a quantum version of the model has suggested the presence of a chiral universality class as well\cite{Chepiga2021}.  In this article, we address the $p=4$ case by studying the classical version of the chiral Ashkin-Teller model, and we show that both possibilities (chiral transition or two-step transition) are realized depending on how close one is to the Potts or clock limit of the model.

The paper is organised as follows. In Section II, we describe the model and the different possibilities for the phase diagram. In Section III, we describe the methodology and the CTMRG algorithm. In Section IV, we benchmark the algorithm and discuss its power and limitation. In Section V we present our results. We compare and study in detail the phase diagram of the chiral clock-Ising model, where a floating phase opens right away, and of the four-state chiral Potts model, where we observe a chiral transition, and we propose a full phase diagram of the model. 

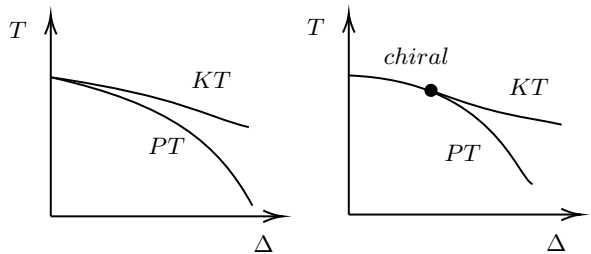
\begin{figure}
\begin{center}
\tikzset{every picture/.style={line width=0.75pt}} %set default line width to 0.75pt        
\begin{tikzpicture}[x=0.75pt,y=0.75pt,yscale=-.9,xscale=.9]
\draw    (99,141) -- (99,30) ;
\draw [shift={(99,28)}, rotate = 450] [color={rgb, 255:red, 0; green, 0; blue, 0 }  ][line width=0.75]    (10.93,-3.29) .. controls (6.95,-1.4) and (3.31,-0.3) .. (0,0) .. controls (3.31,0.3) and (6.95,1.4) .. (10.93,3.29)   ;
\draw    (99,141) -- (227,141) ;
\draw [shift={(229,141)}, rotate = 180] [color={rgb, 255:red, 0; green, 0; blue, 0 }  ][line width=0.75]    (10.93,-3.29) .. controls (6.95,-1.4) and (3.31,-0.3) .. (0,0) .. controls (3.31,0.3) and (6.95,1.4) .. (10.93,3.29)   ;
\draw    (99,63) .. controls (166,78) and (193,100) .. (212,135) ;
\draw    (99,63) .. controls (173,74) and (185,85) .. (210,91) ;
\draw    (266,140) -- (266,29) ;
\draw [shift={(266,27)}, rotate = 450] [color={rgb, 255:red, 0; green, 0; blue, 0 }  ][line width=0.75]    (10.93,-3.29) .. controls (6.95,-1.4) and (3.31,-0.3) .. (0,0) .. controls (3.31,0.3) and (6.95,1.4) .. (10.93,3.29)   ;
\draw    (266,62) .. controls (347,65) and (360,119) .. (369,123) ;
\draw    (312.06,70.36) .. controls (350.36,84.94) and (362.3,84.58) .. (385.3,89.58) ;
\draw  [fill={rgb, 255:red, 0; green, 0; blue, 0 }  ,fill opacity=1 ] (310,68) .. controls (311.3,66.86) and (313.28,67) .. (314.42,68.3) .. controls (315.55,69.61) and (315.41,71.58) .. (314.11,72.72) .. controls (312.81,73.85) and (310.83,73.72) .. (309.7,72.42) .. controls (308.56,71.11) and (308.7,69.14) .. (310,68) -- cycle ;
\draw    (266,140) -- (394,140) ;
\draw [shift={(396,140)}, rotate = 180] [color={rgb, 255:red, 0; green, 0; blue, 0 }  ][line width=0.75]    (10.93,-3.29) .. controls (6.95,-1.4) and (3.31,-0.3) .. (0,0) .. controls (3.31,0.3) and (6.95,1.4) .. (10.93,3.29)   ;
\draw (211,150.4) node [anchor=north west][inner sep=0.75pt]    {$\Delta $};
\draw (375,149.4) node [anchor=north west][inner sep=0.75pt]    {$\Delta $};
\draw (74,30.4) node [anchor=north west][inner sep=0.75pt]    {$T$};
\draw (241,28.4) node [anchor=north west][inner sep=0.75pt]    {$T$};
\draw (152,95.4) node [anchor=north west][inner sep=0.75pt]    {$PT$};
\draw (176,58.4) node [anchor=north west][inner sep=0.75pt]    {$KT$};
\draw (354,62.4) node [anchor=north west][inner sep=0.75pt]    {$KT$};
\draw (318,99.4) node [anchor=north west][inner sep=0.75pt]    {$PT$};
\draw (284,44.4) node [anchor=north west][inner sep=0.75pt]    {$chiral$};
\end{tikzpicture}
\end{center}
\caption{The two possible phase diagrams for $\lambda< \lambda_{c1}$: either the transition is immediately a two-step one through an intermediate critical floating phase, or it is a direct one in the chiral universality class.}
\label{fig:PFlambda0}
\end{figure}

% AT MODEL
\section{Chiral Ashkin-Teller model}

The 2D Ashkin-Teller model (AT)\cite{AshkinTeller} is defined with two Ising spins $\tau, \sigma \in \{\pm 1\}$ on each site with energy:
\begin{align}
	H_0 = - &  \sum_{\langle i,j\rangle}  \sigma_{i} \sigma_{j}  +  \tau_{i} \tau_{j} + \lambda \sigma_{i}\sigma_{j} \tau_{i} \tau_{j} 
\end{align}
where the sum runs over pairs of nearest neighbors. It has been shown\cite{Kohmoto} that there is an exact critical line known from duality where the correlation length exponent varies continuously from the four-state Potts model ($\lambda= 1$) to the clock model ($\lambda=0)$ equivalent to two decoupled Ising models as:
\begin{align}
\nu & = \frac{1}{2-\frac{\pi}{2}\arccos(-\lambda)^{-1}} 
\end{align}
To our knowledge there is no known theoretical value known for $\bar{\beta}$, but it is believed to be larger than 1, implying that the product $\xi q\rightarrow 0$ along this line\cite{HuseFisher1984}. The chiral Ashkin-Teller model can be defined by adding a chiral perturbation along the $x$-direction:
\begin{align}
 H = H_0 + \Delta \sum_{x,y} (\tau_{x+1,y} \sigma_{x,y} - \sigma_{x+1,y} \tau_{x,y})
\end{align}
%This model was originally introduced by Schultz \cite{schulz} where he shows that the chiral perturbation is irrelevant for $\lambda > \lambda_{c1} \sim 0.9779$. Therefore, if $\lambda>\lambda_{c1}$, for sufficiently small values of $\Delta$ the transition will still belong to the AT universality class. On the other hand, for large enough values of $\Delta$, the C-IC transition is expected to be two-step transition separated by a floating phase. In between two scenarios are possible. Either the unique transition always belongs to the AT universality class, or it shifts to the chiral universality class with $\nu_x = \overline{\beta}$ before reaching the Lifshitz point. The two different scenarios are depicted in Fig. \ref{fig:PFlambda0}. Furthermore, for $\lambda=0$, it is commonly agreed that a floating phase appears right away and the transition is always two-step. Thus, if the chiral transition exists at $\lambda>\lambda_{c1}$, it could very well be also present for a range of $\lambda\in[\lambda_{c2}, \lambda_{c1}]$, with $\lambda_{c2}$ strictly positive. We will restrict our study to positive values of $\Delta$ as the partition function is even with respect to $\Delta$.
This model was originally introduced by Schulz \cite{schulz}, who showed that the chiral perturbation is irrelevant for $\lambda > \lambda_{c_1} \sim 0.9779$. For large enough values of $\Delta$, the C-IC transition is expected to be a two-step transition separated by a floating phase. For $\lambda<\lambda_{c_1}$, for small enough values of $\Delta$ two possibilities arise, either the transition is two-step and the floating phase opens right away or the transition is unique and belong to the chiral universality class with $\nu_x = \bar{\beta}$. The two different scenarios are depicted in Fig. \ref{fig:PFlambda0}.
%the transition is unique before reaching a Lifshitz point, and then opens up to a critical phase. 
On the other hand, for $\lambda > \lambda_{c_1}$, the perturbation is irrelevant, and for small enough values of $\Delta$, the transition is expected to be in the AT universality class. The parameter range between $\lambda_{c_1}$ and the Potts limit is very small, actually too small to be observed numerically as we shall see. For larger $\Delta$, beyond this small range of AT transition, the question of a unique chiral transition or a floating phase remains unchanged. %If such unique transition exists, we separate two types of models, the one for which a chiral melting takes place $\lambda > \lambda_{c_2}$ and the one where the transition is either two-step or AT $\lambda < \lambda_{c_2}$. 

%Therefore, we can separate the phase diagram into two categories, for $\lambda> \lambda_{c_2}$ where a chiral melting takes place and for $\lambda<\lambda_{c2}$ where this is not the case.

%In between two scenarios are possible. Either the unique transition always belongs to the AT universality class, or it shifts to the chiral universality class with $\nu_x = \overline{\beta}$ before reaching the Lifshitz point. The two different scenarios are depicted in Fig. \ref{fig:PFlambda0}. Furthermore, for $\lambda=0$, it is commonly agreed that a floating phase appears right away and the transition is always two-step. Thus, if the chiral transition exists at $\lambda>\lambda_{c1}$, it could very well be also present for a range of $\lambda\in[\lambda_{c2}, \lambda_{c1}]$, with $\lambda_{c2}$ strictly positive. We will restrict our study to positive values of $\Delta$ as the partition function is even with respect to $\Delta$.

For the rest of the paper, we will refer to the $\lambda=1$ case as the four-state chiral Potts model and to the $\lambda = 0$ case as the chiral clock-Ising model in reference to their respective universality classes at $\Delta = 0$.
%check clock or no clock, I would say no clock

\section{Methodology}

We use the same methodology already developed in the previous work on the three-state chiral Potts model \cite{nyckees} where more details are available.

\subsection{CTMRG}

The CTMRG algorithm was first introduced by Okunishi and Nishino\cite{nishino}. It is a numerical method which combines Baxter's corner matrices\cite{BaxterBook} and Steve White's renormalisation group density matrix method\cite{dmrg1}. It is most commonly used for two dimensional quantum systems as a contraction algorithm of wave functions\cite{Corboz}, but has recently shown promising results on classical systems as well\cite{ctmrgcl1,ctmrgcl2,ctmrgcl3}. 

\begin{figure}
\begin{center}
\begin{tikzpicture}
%\fill[line width=2pt,color=yellow] (0,0) circle[radius=0.3] 
%\draw (0,0) node{$C_L$};
\draw (-5, -1) node{$\mathcal{Z}= $};
\draw (-0.9, -1) node{$\approx$};
\draw (0.5, 0.2) node{$\chi$};
\draw (-3.6, -0.2) circle (0.2cm) node{$a$}; \draw (-2.8, -0.2) circle (0.2cm) node{$a$}; \draw (-2, -0.2) circle (0.2cm) node{$a$}; 
\draw (-3.6, -1) circle (0.2cm) node{$a$}; \draw (-2.8, -1) circle (0.2cm) node{$a$}; \draw (-2, -1) circle (0.2cm) node{$a$}; 
\draw (-3.6, -1.8) circle (0.2cm) node{$a$}; \draw (-2.8, -1.8) circle (0.2cm) node{$a$}; \draw (-2, -1.8) circle (0.2cm) node{$a$}; 
\draw  [dashed] (-4.3, -0.2)--(-3.8, -0.2);\draw (-3.4, -0.2) -- (-3, -0.2); \draw (-2.6, -0.2)--(-2.2,-0.2); \draw [dashed] (-1.8,-0.2) -- (-1.3,-0.2);
\draw  [dashed] (-4.3, -1)--(-3.8, -1);\draw (-3.4, -1) -- (-3, -1); \draw (-2.6, -1)--(-2.2,-1); \draw [dashed] (-1.8,-1) -- (-1.3,-1);
\draw  [dashed] (-4.3, -1.8)--(-3.8, -1.8);\draw (-3.4, -1.8) -- (-3, -1.8); \draw (-2.6, -1.8)--(-2.2,-1.8); \draw [dashed] (-1.8,-1.8) -- (-1.3,-1.8);
\draw [dashed] (-3.6, 0)--(-3.6, 0.5); \draw [dashed](-2.8, 0)-- (-2.8, 0.5); \draw [dashed](-2, 0)-- (-2, 0.5);
\draw (-3.6, -0.4)-- (-3.6, -0.8); \draw (-2.8, -0.4)-- (-2.8, -0.8); \draw (-2, -0.4)-- (-2, -0.8);
\draw (-3.6, -1.2)-- (-3.6, -1.6); \draw (-2.8, -1.2)-- (-2.8, -1.6); \draw (-2, -1.2)-- (-2, -1.6);
\draw [dashed] (-3.6, -2)--(-3.6, -2.5); \draw [dashed](-2.8, -2)-- (-2.8, -2.5); \draw [dashed](-2, -2)-- (-2, -2.5);
\draw (0,0) circle (0.3cm) node{$C_1$}; \draw (1,0) circle (0.3cm) node{$T_1$}; \draw (2,0) circle (0.3cm) node{$C_2$};
\draw (0,-1) circle (0.3cm) node{$T_4$}; \draw (1, -1) circle(0.2cm) node{$a$}; \draw (2,-1) circle (0.3cm) node{$T_2$};
\draw (0,-2) circle (0.3cm) node{$C_4$}; \draw (1,-2) circle (0.3cm) node{$T_3$}; \draw (2,-2) circle (0.3cm) node{$C_3$};

\draw [line width=0.3mm] (0.3,0) --  (0.7,0); \draw [line width=0.3mm] (1.3,0) -- (1.7,0);
\draw [line width=0.3mm] (0,-0.3)--(0,-0.7); \draw (1,-0.3) -- (1,-0.8); \draw  [line width=0.3mm] (2,-0.3) -- (2,-0.7);
\draw (0.3,-1) -- (0.8,-1); \draw (1.7,-1) -- (1.2,-1);
\draw [line width=0.3mm] (0,-1.3) -- (0,-1.7);  \draw (1,-1.7) -- (1, -1.2);\draw [line width=0.3mm] (2,-1.3) -- (2,-1.7);
\draw  [line width=0.3mm] (1.7,-2) -- (1.3,-2); \draw   [line width=0.3mm](0.7,-2) -- (0.3,-2);
\end{tikzpicture}
\end{center}
\caption{Sketch of the partition function as the contraction of a tensor network.}
\label{fig:partitionFunction}
\end{figure}
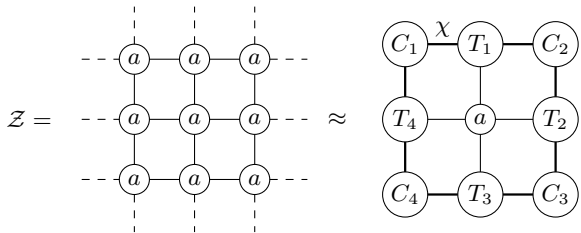

The investigation of classical systems with tensor networks is done by expressing the partition function in the thermodynamic limit as an infinite tensor network and contracting it. CTMRG sets a way to contract such an infinite square tensor network made of local tensors $a$ on each vertex as shown in Fig. \ref{fig:partitionFunction}. In particular, it allows one to express the Gibbs measure observables as a contraction made of eight tensors denoted by $\mathcal{T} = \{C_1, T_1, C_2, T_2, C_3, T_3, C_4, T_4 \}$, with corner tensors $C_i$ of dimension $\chi \times \chi$ and row/ column tensors $T_i$ of dimensions $\chi \times 4 \times \chi$. The parameter $\chi$ is what controls the numerical approximation and is usually referred to as the bond dimension. The local tensors $a$ can be expressed in many different ways, we choose the most common one given by
\begin{align}
a_{i_1 i_2 i_3 i_4} = \sum_{j_1 j_2 j_3 j_4} \sqrt{Q_{i_1 j_1}^x} \sqrt{Q_{j_3 i_3}^x} \sqrt{Q_{i_2 j_2}^y} \sqrt{Q_{j_4 i_4}^y}
\end{align}
with $Q_{ij}^{x/y}$ the Boltzmann weight matrices between two spins on the horizontal/vertical axis. In the $(\sigma, \tau) = \{(1,1), (1,-1),(-1,1),(-1,-1)\}$ basis, $Q^x$ is given by:
\begin{align}
Q^x & = 
\begin{pmatrix}
e^{x_0} & e^{x_1} & e^{x_2} & e^{x_3}\\
e^{x_2} & e^{x_0} & e^{x_3} & e^{x_1}\\
e^{x_1} & e^{x_3} & e^{x_0} & e^{x_2}\\
e^{x_3} & e^{x_2} & e^{x_1} & e^{x_0}\\
\end{pmatrix}
\end{align}
with 
\begin{align*}
&x_0 = \beta(\lambda+2), \quad x_1 = -\beta( 2\Delta  + \lambda )\\
&x_3 = \beta(\lambda-2), \quad x_2 = \beta(2\Delta-\lambda))
\end{align*}
$Q^y$ is defined similarly with $\Delta = 0$.

CTMRG therefore provides an easy way to measure directly local observable and two points functions. % as shown in Fig. \ref{}. 

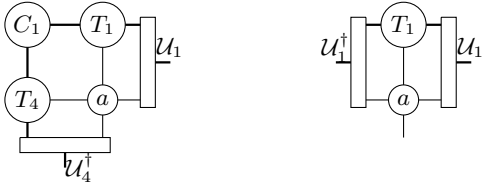
\begin{figure}
\begin{center}
\begin{tikzpicture}
\draw (0,0) circle (0.3cm) node{$C_1$}; \draw (1,0) circle (0.3cm) node{$T_1$};
\draw (0,-1) circle (0.3cm) node{$T_4$}; \draw (1,-1) circle (0.2cm) node{$a$};
\draw [line width=0.3mm] (0.3, 0) -- (0.7,0); \draw  [line width=0.3mm](1.3,0)--(1.5,0);
\draw [line width=0.3mm] (0,-0.3) -- (0,-0.7); \draw (1,-0.3)--(1, -0.8);
\draw  (0.3, -1) -- (0.8,-1); \draw (1.2,-1)--(1.5,-1);
\draw [line width=0.3mm] (0,-1.3)-- (0,-1.5); \draw (1,-1.2)--(1,-1.5);
\draw (1.5,0.1) -- (1.5,-1.1) -- (1.7,-1.1) -- (1.7,0.1) -- (1.5,0.1);
\draw [line width=0.3mm] (1.7, -0.5)--(1.9, -0.5);
\draw (1.9, -0.3) node{$\mathcal{U}_1$};
\draw (-0.1,-1.5) -- (-0.1,-1.7) -- (1.1,-1.7) -- (1.1,-1.5) -- (-0.1,-1.5);
\draw [line width=0.3mm] (0.5, -1.7)--(0.5, -1.9);
\draw (0.7, -1.9) node{$\mathcal{U}_{4}^{\dagger}$};

\draw [line width=0.3mm] (4.5,0)--(4.7,0);  \draw (5,0) circle (0.3cm) node{$T_1$}; \draw [line width=0.3mm] (5.3,0)--(5.5,0); 
\draw (5,-0.3)--(5,-0.8);
\draw (4.5,-1)--(4.8,-1);  \draw (5,-1) circle (0.2cm) node{$a$}; \draw (5.2,-1)--(5.5,-1); 
\draw (5,-1.2)--(5,-1.5);
\draw (5.5, 0.1) -- (5.5,-1.1) -- (5.7, -1.1) -- (5.7,0.1) -- (5.5, 0.1);
\draw [line width=0.3mm] (5.7,-0.5)--(5.9,-0.5);
\draw (5.9, -0.3) node{$\mathcal{U}_1$};
\draw (4.5, 0.1) -- (4.3,0.1) -- (4.3, -1.1) -- (4.5,-1.1) -- (4.5, 0.1);
\draw [line width=0.3mm] (4.1,-0.5)--(4.3,-0.5);
\draw (4.1, -0.3) node{$\mathcal{U}_{1}^{\dagger}$};
\end{tikzpicture}
\caption{Sketch of the full iteration for the corner and row tensors $C_1$ and $T_1$.}
\label{fig:iteration}
\end{center}
\end{figure}

The algorithm can be summarised in two steps \cite{orus}: 

(i) \textbf{Update}: each corner matrix is extended by adding a row, a column and a local $a$ tensor. And each row/ column tensor is extended by adding a local tensor $a$. We illustrate this step for the tensors $C_1$ and $T_1$ in Fig. \ref{fig:iteration}. The other tensors are updated similarly.

(ii)\textbf{Truncation:} the update increases the bond dimension of the environment tensors by a factor 4. In order to avoid an exponentially growing bond dimension one needs to project the tensors into a subspace.  The isometries are given by doing the singular value decomposition on some density matrices and keeping only the $\chi$ largest singular values. The choice of the density matrices influences greatly the convergence of the algorithm. We choose the ones originally proposed by Nishino and Oknishi and given by:
\begin{align}
&\mathcal{U}_1' \mathcal{S}_1' \mathcal{V}_1' = C_2' C_3' C_4' C_1' \\
& \mathcal{U}_2' \mathcal{S}_2' \mathcal{V}_2' = C_3' C_4' C_1' C_2' \\
& \mathcal{U}_3' \mathcal{S}_3' \mathcal{V}_3' = C_4' C_1' C_2' C_3' \\
& \mathcal{U}_4' \mathcal{S}_4' \mathcal{V}_4' = C_1' C_2' C_3' C_4'  
\end{align}
with $C_i'$ the extended corner matrices. A full iteration for one corner and one column tensor is illustrated in Fig. \ref{fig:iteration}.

Both steps are repeated until the energy between two iterations converges to some threshold, at which point we consider the thermodynamic limit reached. We note that the initial tensors represent the boundary conditions on the infinite systems and influence the convergence of the algorithm. We choose to set the boundary conditions to be open as it gives the best convergence.

%EFFECTIVE EXPONENT
\subsection{Effective exponents and critical temperatures}
Our analysis is based on effective exponents and some general hypotheses regarding the nature of the transitions. The effective exponent of some physical quantity $A$ which behaves algebraically $A\sim t^{\theta}$ is defined as:
\begin{align}
\theta_\text{eff} = \frac{d\log(A)}{d\log(t)}
\end{align}
with $t = \mid T -T_c\mid$, the reduced temperature. Due to crossovers and various corrections, the effective exponent might not be equal to the critical exponent if evaluated outside the critical regime but should converge to it as $t\rightarrow 0$.  If the C-IC transition is unique, effective exponents, and in particular $\nu_\text{eff}$ from both sides of the transition, should converge to the same value at $t=0$. It turns out that this requirement is enough to fix $T_c$. On the other hand, if the transition is two-step, due to the exponential divergence of the correlation length from the KT transition at high temperatures, asking for a unique limit of $\nu_\text{eff}$ from both sides of the transition will lead to $\nu>1$. Such an exponent is unphysical and one can conclude that the hypothesis is wrong. Thus, the transition must be two-step. In that case, fixing $T_{KT}$ by looking for the regime where $\eta=1/4$ turned out to be more accurate than fitting the correlation with an exponential, where $\eta$ is the exponent of the decay of the spin-spin correlation function:
\begin{align}
 \langle S_0 \cdot S_r\rangle = \langle\tau_0\tau_r\rangle + \langle \sigma_0\sigma_r\rangle \propto r^{-\eta}
\end{align}
$T_{PT}$ is fixed by the condition $\lim_{t\rightarrow 0} \nu_\text{eff}^y =1$ from the ordered phase. From now on, we will use $\nu^{HT}$ and $\nu^{LT}$ to denote the limit of the effective exponent from the respective high and low temperature regimes.

%% EXTRAPOLATION
\subsection{Extrapolation of the correlation length and wave-vector}

\begin{figure}[t!]
\includegraphics[width=0.45\textwidth]{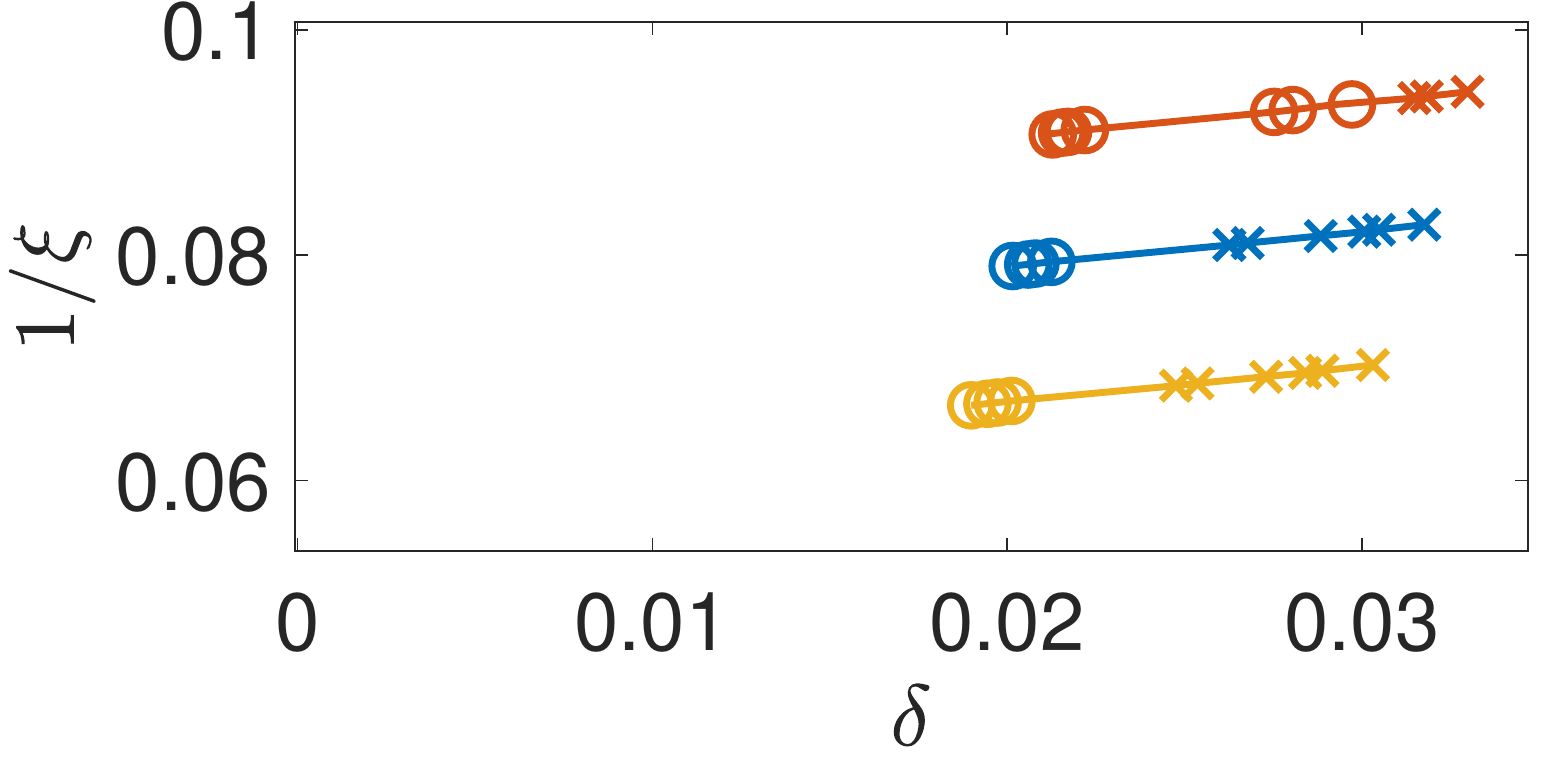}
\caption{Inverse correlation length as function of a gap of the transfer matrix at $\Delta = 0.2$ and $\lambda = 1$. The bond dimensions used vary from $100$ to $200$. Because of a level crossing, two different eigenvalues were used to extrapolate the correlation in order to follow the same gap : $\delta = \epsilon_{18}- \epsilon_5$ (crosses) and  $\delta = \epsilon_{16} - \epsilon_5$ (circles). Each color represents a different temperature. The simulations are done in the low temperature regime.}% the larger the correlation length extrapolates too, the closer the temperature is to $T_c$.}
\label{fig:extrapolation}
\end{figure}

The correlation length and wave-vector can be easily computed through partial diagonalisation of transfer matrices. If one denotes the normalised ordered eigenvalues of the transfer matrix with $\lambda_i = \epsilon_j e^{i\phi_j}$, $j=1,2,\ldots$, then the correlation length and wave-vector are given by:
\begin{align}
 \xi & = \frac{1}{\epsilon_2},  \qquad q = \phi_2
\end{align}
This estimation, which corresponds to a given finite bond dimension, can be improved using extrapolations\cite{czarnik2018} with respect to gaps ($\delta)$ within the transfer matrix spectrum. Rams et al suggested that the inverse correlation length scales linearly such that:
\begin{align}
& \frac{1}{\xi_{\text{exact}}} = \epsilon_2 + \delta \\
& \phi_{\text{exact}} = \phi_2 + \delta' 
\end{align}
Various gaps $\delta$ can be used. In particular, in the disordered phase we systematically used $\delta = \epsilon_4 -\epsilon_2$. In some cases the energy levels of the transfer matrices cross and in order to follow a given gap closing, we extrapolate using different eigenvalues. An example of such a case is given in Fig. \ref{fig:extrapolation}.
%We noticed that, in some cases, the gaps converge to some finite values, and one need to look at higher ones to obtain correct extrapolations. An example of such a case is given in Fig. \ref{fig:extrapolation}. We note that the spectrum of the transfer matrix does not need to be continuous unless the exponentially decaying connected correlations have an algebraic pre-factor. In Fig. \ref{fig:extrapolation}, the gaps converge with respect to $\chi$ to finite values for the lowest temperatures. Thus it seems that the correlations are purely exponential when the temperature is too far from criticality. 
As for the wave-vector extrapolation, we used differences in the argument of the eigenvalues, usually we used $\delta' = \phi_4 -\phi_2$.

%ERROR BARS
\subsection{Error bars}
The errors on the effective exponents come from the uncertainty on the critical temperature $\delta T_c$ and the error in the correlation length from the linear fit $\delta \xi$. A first order Taylor expansion gives: 
\begin{align*}
\delta \nu_\text{eff}= & \nu_\text{eff} \left(  \frac{\delta \xi(T)}{\xi(T)} + \frac{\delta T_c}{T -T_c} \right. \\
	& \left. + \frac{\delta\xi(T+dT)+ \delta\xi(T-dT)}{\xi(T+dT)- \xi(T-dT)} \right)
\end{align*}
Error bars on the critical exponents are computed by fitting error bars of the effective exponents as shown in Fig.\ref{fig:IsingBM}

% BENCHMARK
\section{Benchmark}

In this section we benchmark the results on the two extreme cases $\lambda = 0$ and $\lambda = 1$ of the well understood Ashkin-Teller line in order to test and understand the limitations of the algorithm. Both models are commensurate along $\Delta = 0$, so in order to study $\bar{\beta}$ we approach the critical point from the line $\Delta =T - T_c$.

\subsubsection{Ising point}

At the $\lambda = \Delta = 0$ point, the model decouples into two independent Ising models. Therefore, the transition belongs to the Ising universality class characterised by an exponent $\nu = 1$ and a specific heat that diverges logarithmically, i.e. an exponent $\alpha = 0$ with a multiplicative logarithmic correction. The critical temperature is known from duality and is given by $T_c = 2/\log(\sqrt{2}+1)$. 

\begin{figure}[t!]
\includegraphics[width=0.45\textwidth]{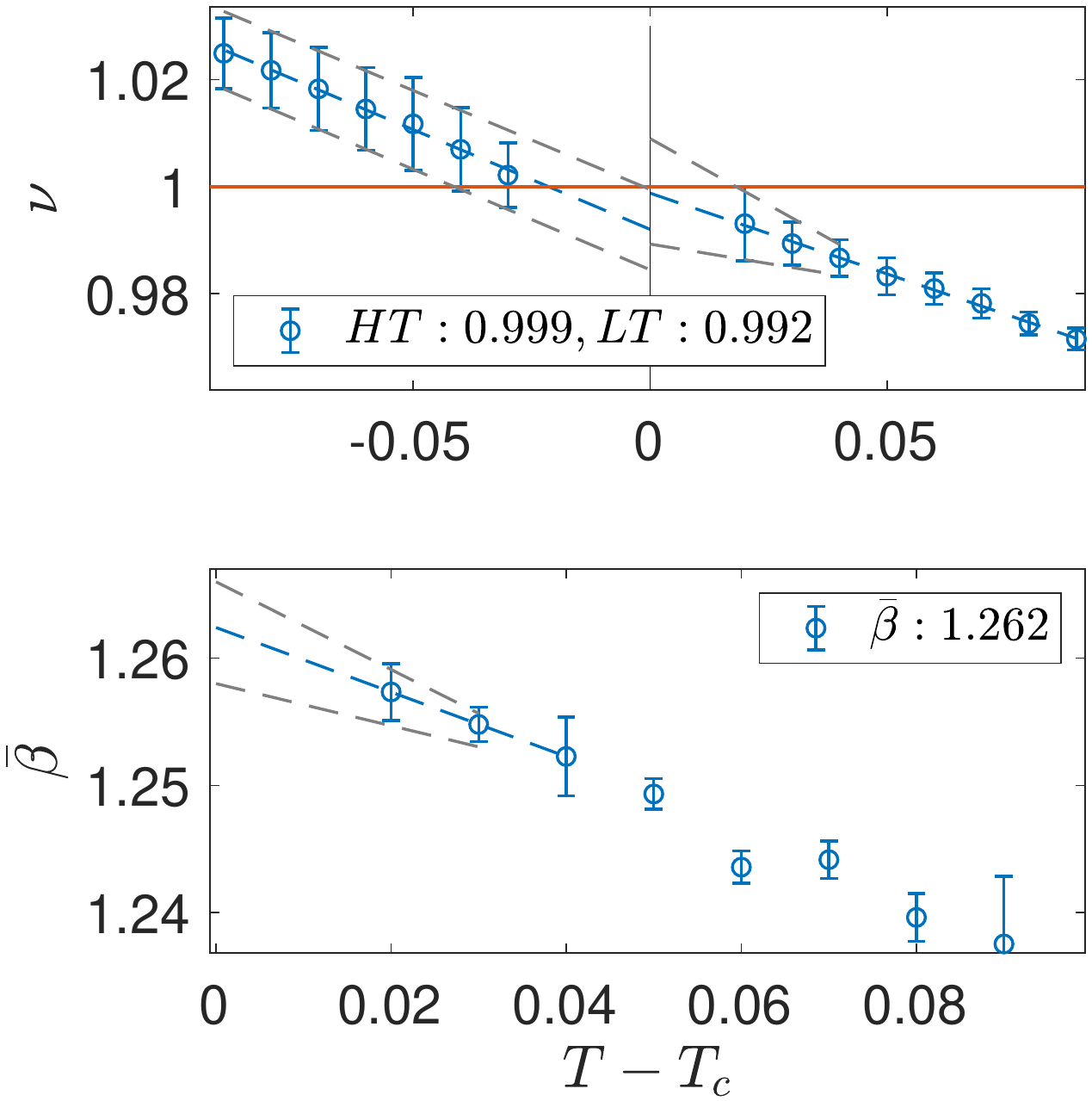}
\caption{Effective exponents for the Ising model. We extrapolated with a linear fit $\nu= 0.992\pm 0.008$ and $\nu = 0.999\pm 0.01$ respectively from the low and high temperature regime. The error on the extrapolated values is estimated with a linear fit (grey dashed line) on the error bars of the effective exponents. For $\bar{\beta}$, a linear fit on the last three points extrapolates to $\bar{\beta} = 1.262\pm 0.004$.}
\label{fig:IsingBM}
\end{figure}

The numerical results are summarised in Fig. \ref{fig:IsingBM}. A linear extrapolation gives $\nu^{HT} = 0.999\pm 0.01$ and $\nu^{LT}=0.992\pm 0.008$ in excellent agreement with the theoretical value $\nu = 1$. The wave-vector exponent extrapolates to $\bar{\beta}= 1.262\pm 0.004$, implying that the product $\xi q\rightarrow 0$.

\subsubsection{Four-state Potts point}

We now turn to the $\Delta = 0$ and $\lambda = 1$ point, which belongs to the four-State Potts universality class. It is characterised by the exponents $\nu = 2/3$ and $\alpha = 2/3$ with multiplicative logarithm corrections\cite{Salas1997} both in the specific heat and in the correlation length. The critical temperature $T_c = 4/\log 3$ is also known from duality .

\begin{figure}[t!]
\includegraphics[width=0.45\textwidth]{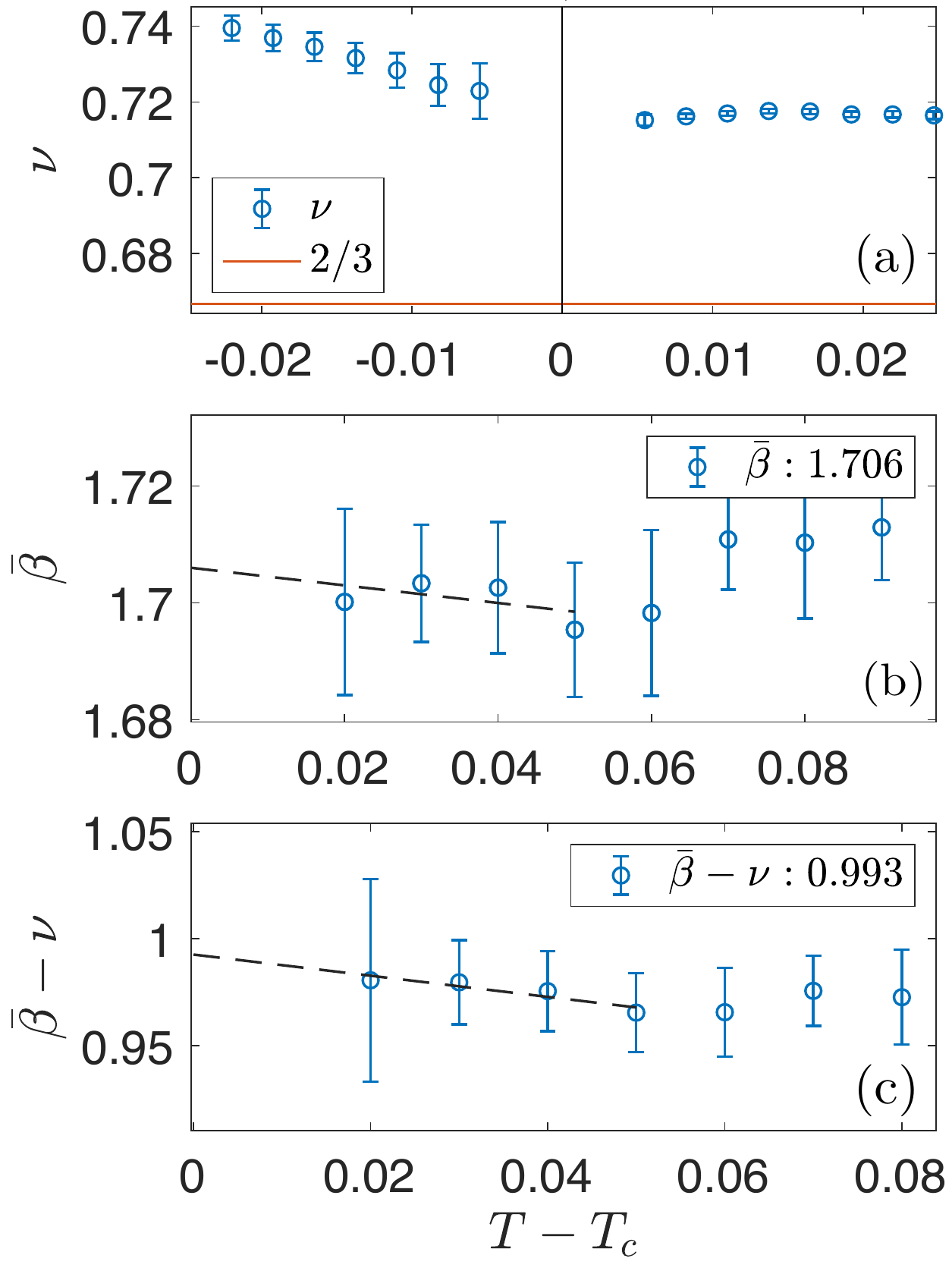}
\caption{Effective exponents for the Potts model. From top to bottom: (a) Simulations done along the commensurate line; (b)\&(c) Simulations done along the line $\Delta \propto T -T_c$. }
\label{fig:PottsBM}
\end{figure}

Due to these corrections, the measurement of $\nu$ is known to be difficult. In particular, previous studies with the Monte Carlo Renormalization group method\cite{Eschbach} have obtained  $\nu^{-1}\simeq1.34$ (or $\nu\simeq 0.746$).% and $\nu=0.752$\cite{Rebbi}. 
 We find similar results. A linear extrapolations of the effective exponent on Fig.\ref{fig:PottsBM} gives $\nu = 0.713$. This is quite far from the exact value. Indeed, due to the logarithmic correction, extrapolating the results with a linear fit is wrong and one expects the effective exponent to converge to 2/3 with an infinite slope. We are not close enough to the critical temperature to observe this change of behaviour.
For the incommensurability, we observe $\bar{\beta} = 1.71\pm 0.02$. Due to the poor accuracy of $\nu$, we can ask how much one can trust this result. However, we notice that the product $\xi q$ has exponent $\bar{\beta} -\nu = 0.99 \pm 0.06$. Thus, if the product cancels the corrections in the wave-vector and in the correlation length, one would have $\bar{\beta} = 2/3+0.99 = 1.66 \pm0.06$ which we note to be close to the three-state Potts critical exponent. Results are summarised in Fig. \ref{fig:PottsBM}. In any case, we established that, as expected $\bar{\beta} > \nu $ and $\xi q\rightarrow 0$.

% beta : -0.1504*x + 1.706 \in[1.689, 1.722]	  -> 1.70
% nu: .3439*x + 0.713 \in [0.7116 0.7145] 		  -> 0.713 [0.712, 0.714]
% beta - nu :  -.4943*x + 0.9926 \in [0.9359 1.049] - > 0.99 \in [0.94, 1.05]

\section{Results for $0\leq \lambda \leq 1$}

We now look at the C-IC transitions. In particular, we study in detail the phase diagram of the chiral Ising and the four-state chiral Potts models. For both systems, we observe a KT and a PT transition enclosing a floating phase. In addition, for the four-state chiral Potts model, we found a unique chiral transition.

\subsection{Pokrovsky-Talapov transition}

\begin{figure}[t!]
\includegraphics[width=0.45\textwidth]{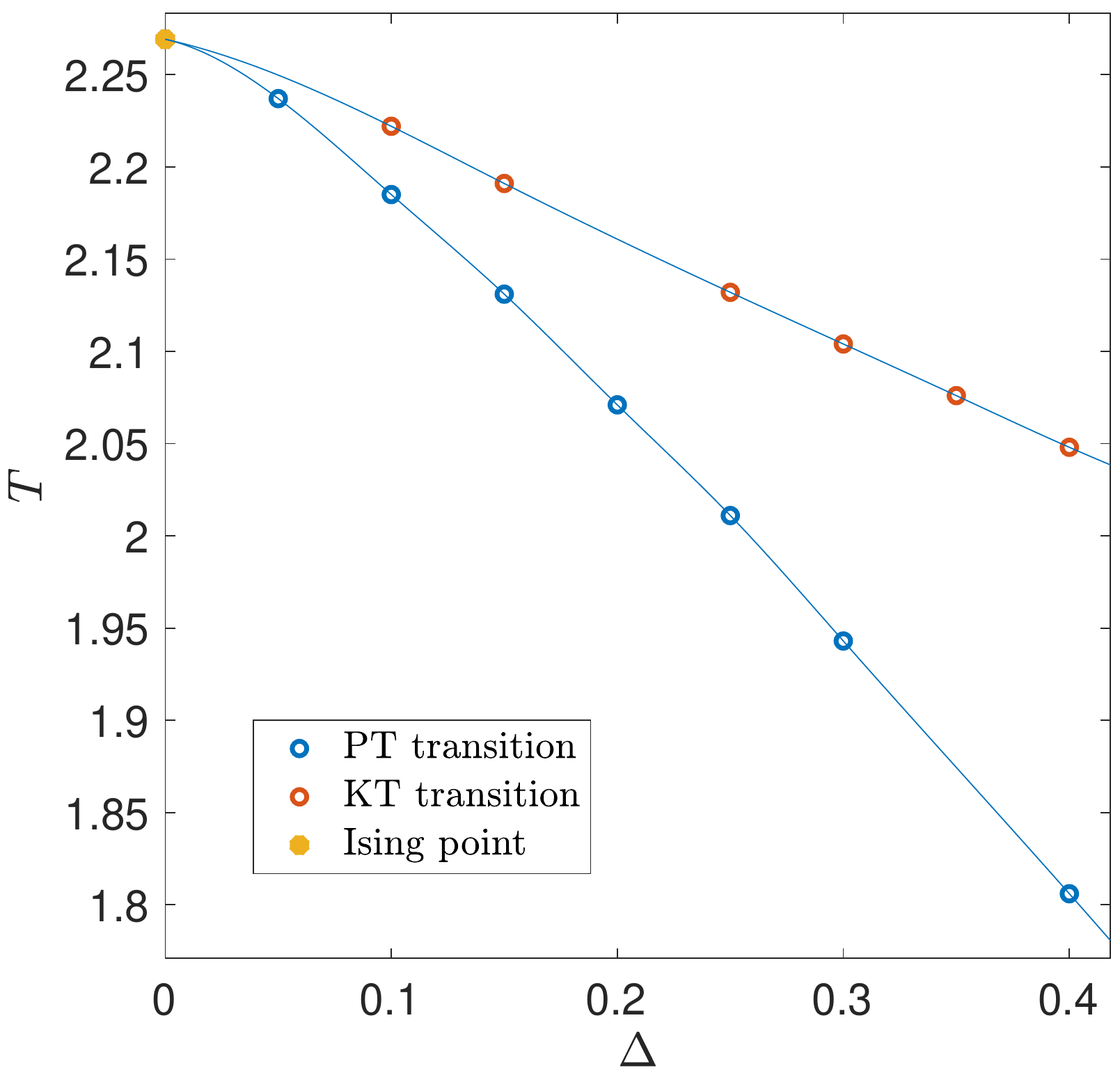}
\caption{Phase diagram for $\lambda = 0$. The red and blue dots represent the measured critical temperatures. The lines are spline interpolations. }
\label{fig:PDIsing}
\end{figure}

We first consider the $\lambda = 0$ model at which we observe a PT transition for $\Delta$ as small as 0.05. As we will see later, in contrast to larger values of $\lambda$, the floating phase extends over a broad parameter range for relatively small values of $\Delta$ and we are able to locate $T_{KT}$ using the criterion $\eta = 1/4$ for $\Delta >0.1$. Looking at the phase diagram of Fig. \ref{fig:PDIsing}, the KT and the PT transitions do not seem to cross before the Ising point and we conclude that, as expected for this model, the floating phase opens up as soon as the chirality is introduced.

\begin{figure}[t!]
\includegraphics[width=0.45\textwidth]{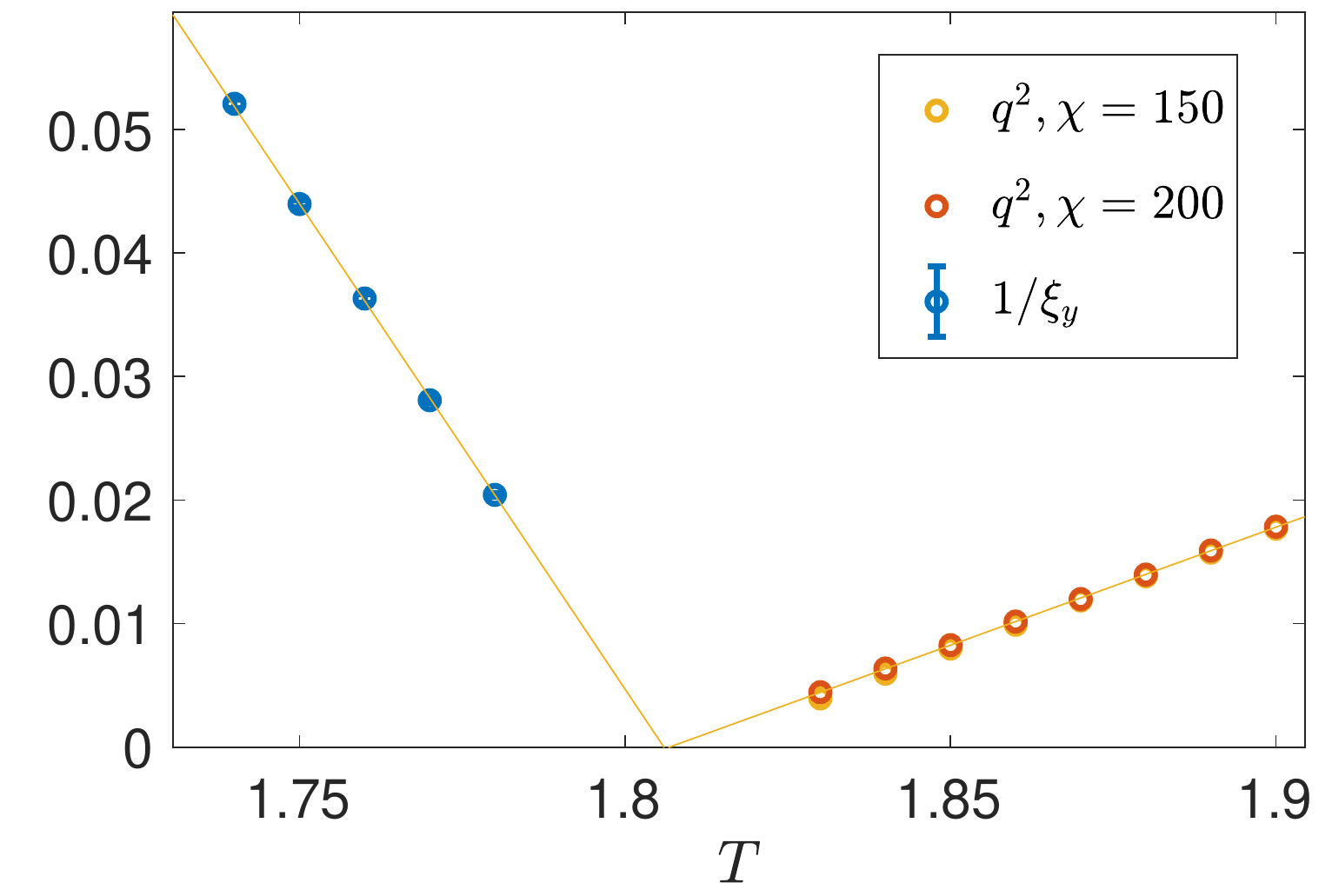}
\caption{Inverse correlation length and square of the wave-vector as a function of temperature at $\Delta = 0.4$ and $\lambda = 0$. If $\nu_y = 1$, $T_{PT}$ can be determined by the intersection of $1/\xi_y$ and the temperature axis. If $\bar{\beta}=1/2$, the same applies to the squared wave-vector. The critical temperature extracted from the correlation length gives $T_{PT}= 1.806$, and from the wave-vector $T_{PT}=1.807$, both in good agreement with each other. We also plot the squared wave-vector obtained for a smaller bond dimension.}
\label{fig:PTIsing}
\end{figure}

\begin{figure}[t!]
\includegraphics[width=0.47\textwidth]{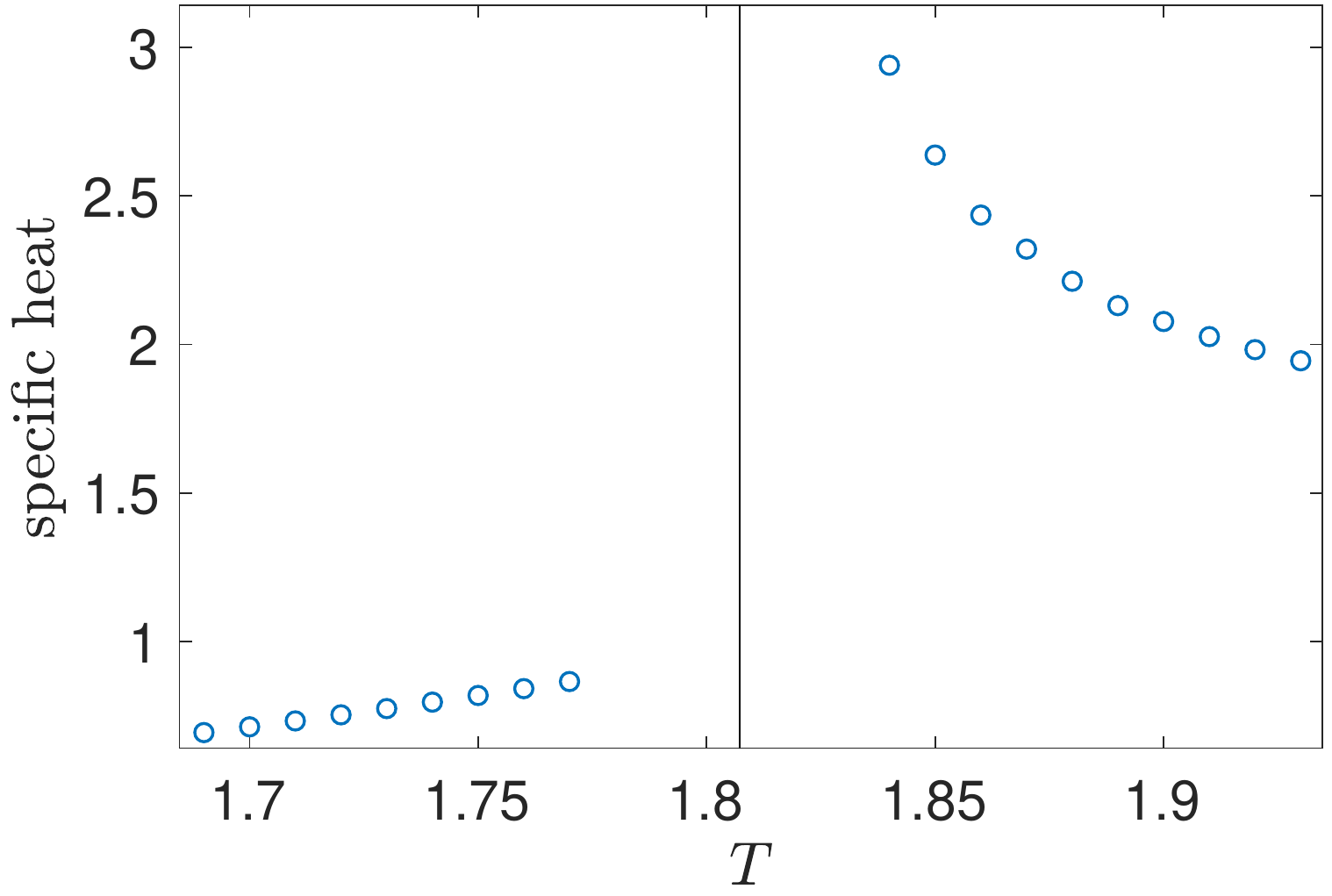}
\caption{Specific heat at $\Delta = 0.4$ and $\lambda = 0$ for finite bond dimension, $\chi = 200$ at low temperature and $\chi = 150$ at high temperature. As there is no extrapolation over the bond dimension, we prefer not to use effective exponents. The specific heat does not diverge from the ordered phase and diverges from the floating phase, as expected for a PT transition. }
\label{fig:specHeatPT}
\end{figure}

In Fig.\ref{fig:PTIsing}, we present a more careful study of the transition at $\Delta = 0.4$. We have plotted the inverse correlation length in the $y$-direction and the square of the wave vector with respect to the temperature. We were not able to measure $\xi_x$ from the ordered phase due to poor extrapolations. Because in the floating phase the algorithm converges very slowly, the wave-vector values shown have not been extrapolated but were made with fixed bond dimension $\chi = 200$ and $\chi = 150$. One first notice the accuracy of both linear fits on a large temperature interval, in agreement with critical exponent $\nu_y = 1$ and $\bar{\beta}=1/2$ with no or very little corrections. If these are indeed the correct critical exponents, the critical temperature can be determined by the intersection with the temperature axis either using the inverse correlation lenght or with the squared wave-vector. This gives respective values of $T_{PT} = 1.806$ and $T_{PT} = 1.807$, in agreement with each other up to $10^{-3}$, and provides a self consistent picture in agreement with a PT transition. Furthermore, in Fig.~\ref{fig:specHeatPT}, we observe that the specific heat does not diverge from the ordered phase but has a clear divergence from the floating phase, as expected at a PT transition. 

For $\lambda = 1$, we observe a PT transition only at $\Delta >0.7$ with at first, a narrow floating phase. We illustrate the results at $\Delta = 0.8$ in Fig. \ref{fig:PDCP3}. For $\lambda =0$ and $\Delta = 0.4$, we deduced $T_{PT}$ from the critical exponents. Here, we work the other way around, we fix $T_{PT}$ and deduce the exponents. Setting $T_{PT}$ with $\nu_{y}^{LT}=1$ at the transition fixes $\bar{\beta}= 0.53\pm0.04$ and $\nu_x^{LT} = 0.51\pm 0.05$, once again providing a self consistent picture with exponents in good agreement with the PT universality class. 

\begin{figure}[t!]
\includegraphics[width=0.45\textwidth]{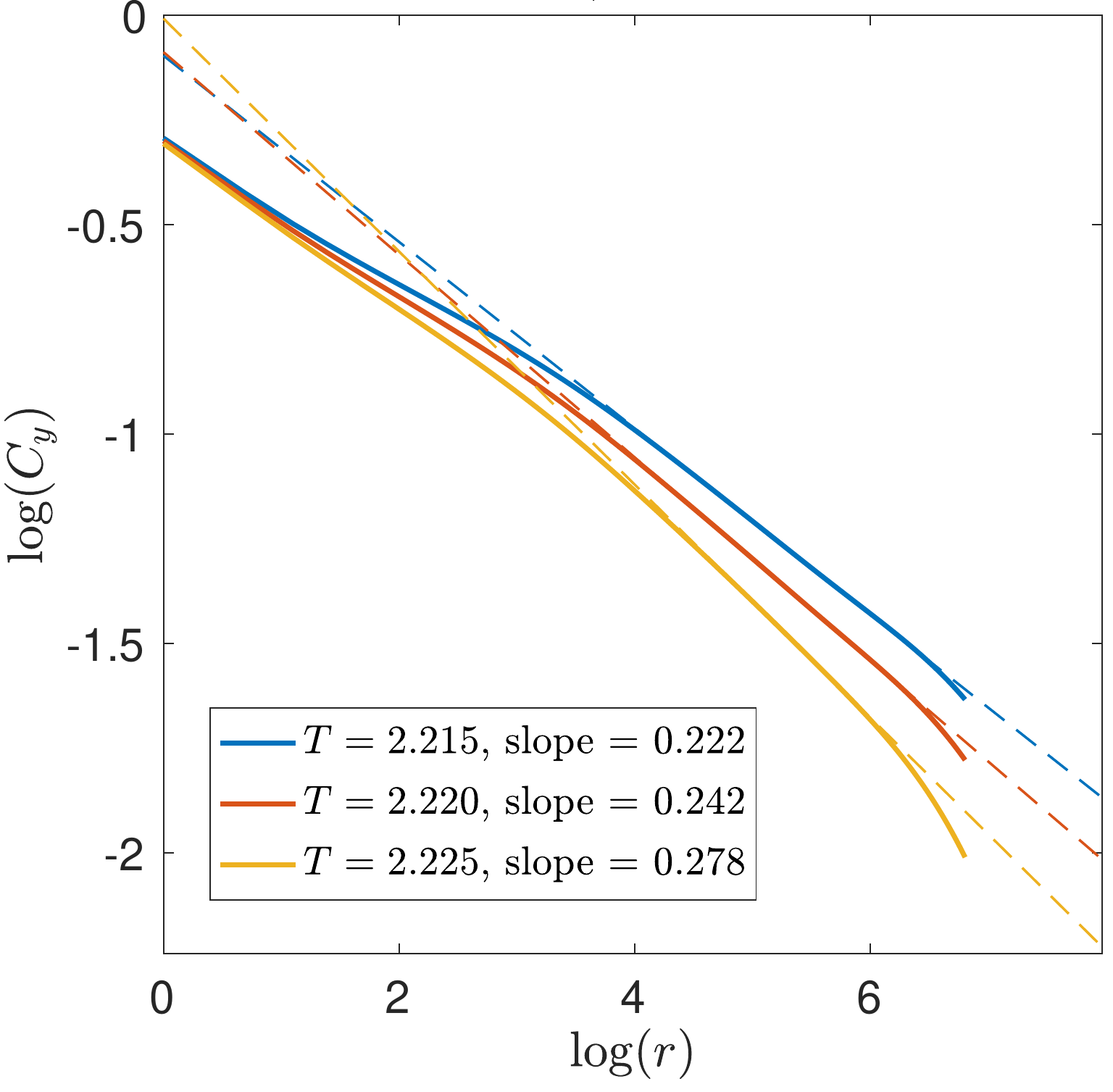}
\caption{Log-log plot of the spin-spin correlations for $\Delta = 0.1$ at $\lambda = 0$. The bond dimension used is 200 and the convergence in energy $10^{-9}$. The results are similar up to 400 sites for bond dimension 150. The fits are done between 150 and 300 sites. The exponent reaches the values $\eta = 1/4$ between $T = 2.22$ and $T = 2.225$, allowing us to locate $T_{KT}$ with a precision of the order $5 \cdot 10^{-3}$. In general, the larger the floating phase, the slower the exponent $\eta$ varies inside it. Therefore, we have best estimates of the critical temperature for narrow floating phases rather than larger ones.}
\label{fig:loglogKT}
\end{figure}

\begin{figure}[t!]
\includegraphics[width=0.45\textwidth]{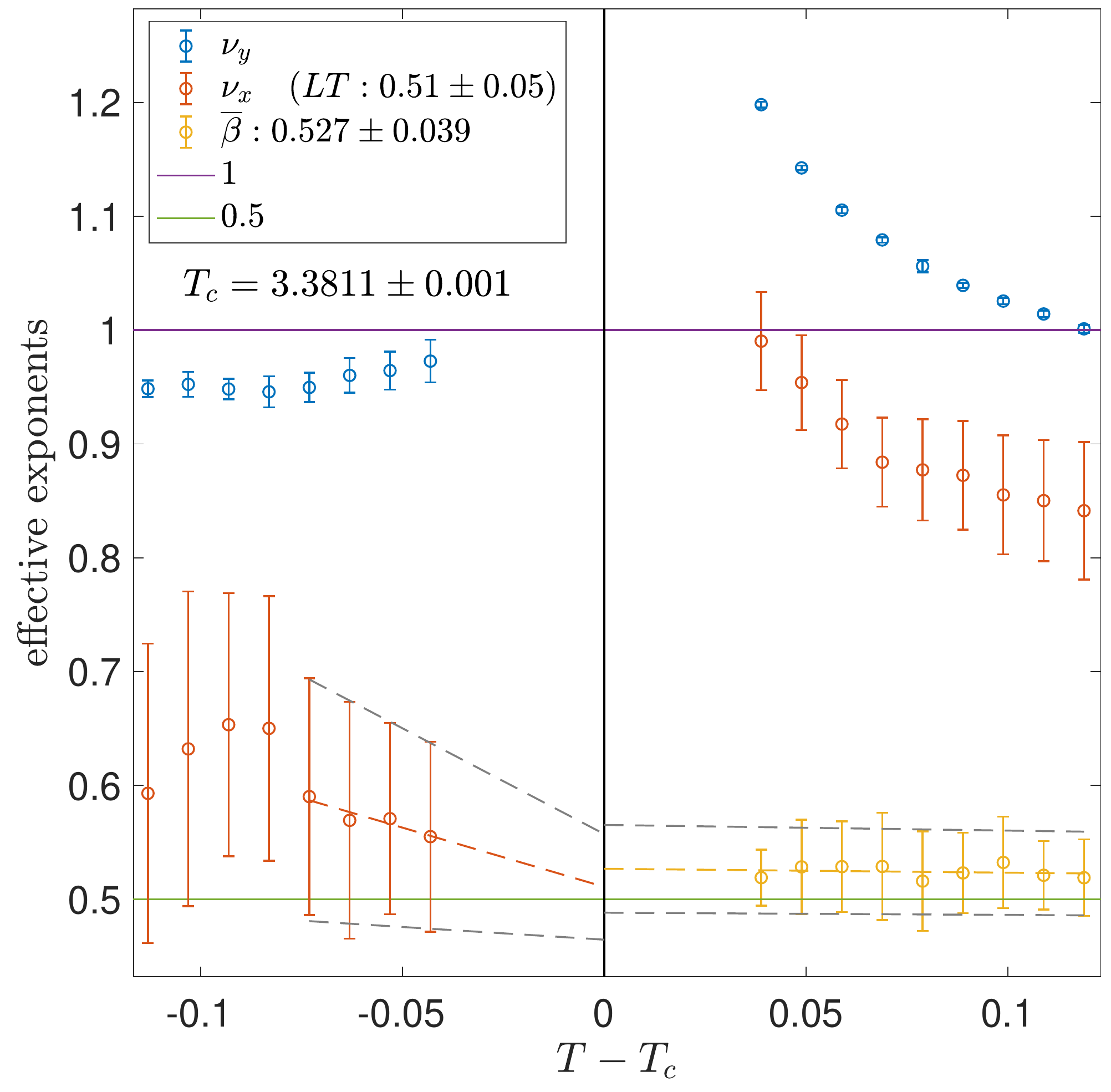}
\caption{Effective exponents as function of temperature for $\Delta = 0.8$ at $\lambda = 1$. $\nu_x$ and $\nu_x$ diverge from the disordered phase in agreement with an exponential divergence of the correlation length. We found $\bar{\beta}= 0.53 \pm 0.04$ and $\nu_x^{LT} = 0.51\pm 0.05$. $\bar{\beta}$ is extrapolated with a linear fit on all available points. $\nu_x$ is extrapolated with a linear fit on the four last points. The error is computed by fitting the error bars and indicated with a grey dashed line. }\label{fig:PDCP3}
\end{figure}
% 1.0406*x + 0.5108
% -1.8584 * x + 0.55718
%  -0.2228* x + 0.4644

\subsection{Kosterlitz-Thouless transition}

The main limitation of the algorithm is its convergence. It turns out that close to the PT transition, the algorithm does not converge to sufficiently low precision to give sensible results. This forbids us to use the $\eta = 1/4$ criteria to determine $T_{KT}$ when the floating phase is too narrow. In that case, an exponential fit of the correlation length will not give enough accuracy to distinguish the two critical temperatures. 

As already mentioned, for $\lambda = 0$, the floating phase extends over a large parameter range so that even at $\Delta = 0.1$ the $\eta=1/4$ criterion can be used to locate the KT transition. We now illustrate such log-log fit with different temperatures in Fig. \ref{fig:loglogKT}. We notice that the log-log plot is linear only above some threshold distance that increases as the temperature decreases. This is due to the fact that the true power law decay is only visible above the average distance between domain walls $l= \frac{2\pi}{4q}$ which diverges as one approaches the PT transition. For $T = 2.22$, we evaluate $\log l \simeq 3.99$ which is in good quantitative agreement of where the curve starts to be linear in Fig. \ref{fig:loglogKT}. This makes the study of the power law decay close to the transition extremely difficult. 

For $\lambda = 1$, we can use the $\eta =1/4$ criterion to determine $T_{KT}$ only above $\Delta>1$. In contrast, for $0.7<\Delta<1$ the floating phase is too narrow to distinguish the two critical temperature transitions. In any case we can identify the transition to be in the KT universality class from the divergence of $\nu_x$ and $\nu_y$ from the high temperature regime at the transition as shown in Fig. \ref{fig:PDCP3}.

\subsection{Chiral transition and Lifshitz point}

\begin{figure}[t!]
\includegraphics[width=0.45\textwidth]{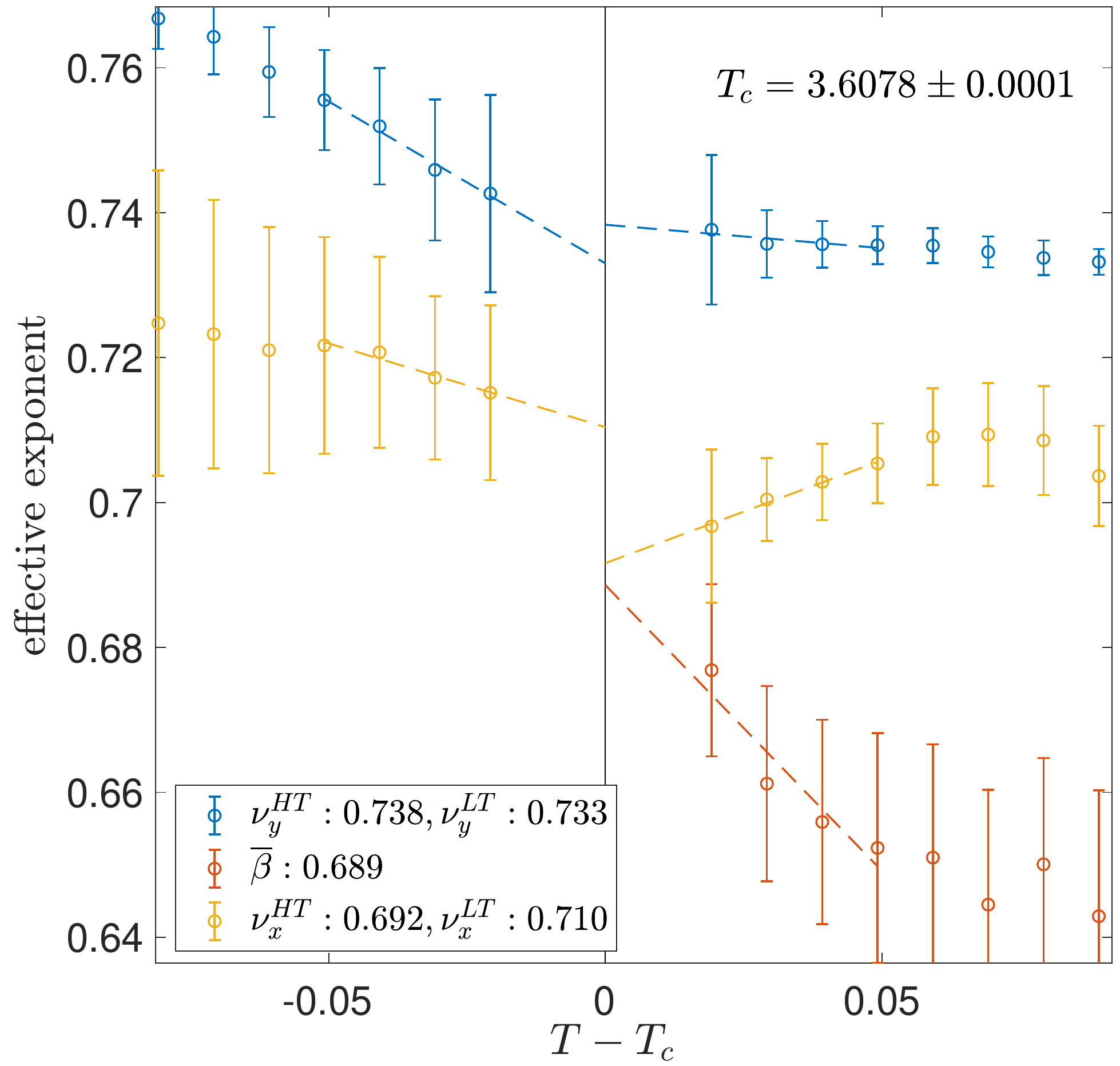}
\caption{Effective exponents as a function of temperature for $\Delta = 0.3$ at $\lambda = 1$. Simulations were made with bond dimension between 100 and 200. The main part of the error bars is due to the extrapolations error. Extrapolated exponents come from a linear fit done on the four last points (dashed lined). }
\label{fig:Chiral03}
\end{figure}

\begin{figure}[t!]
\includegraphics[width=0.45\textwidth]{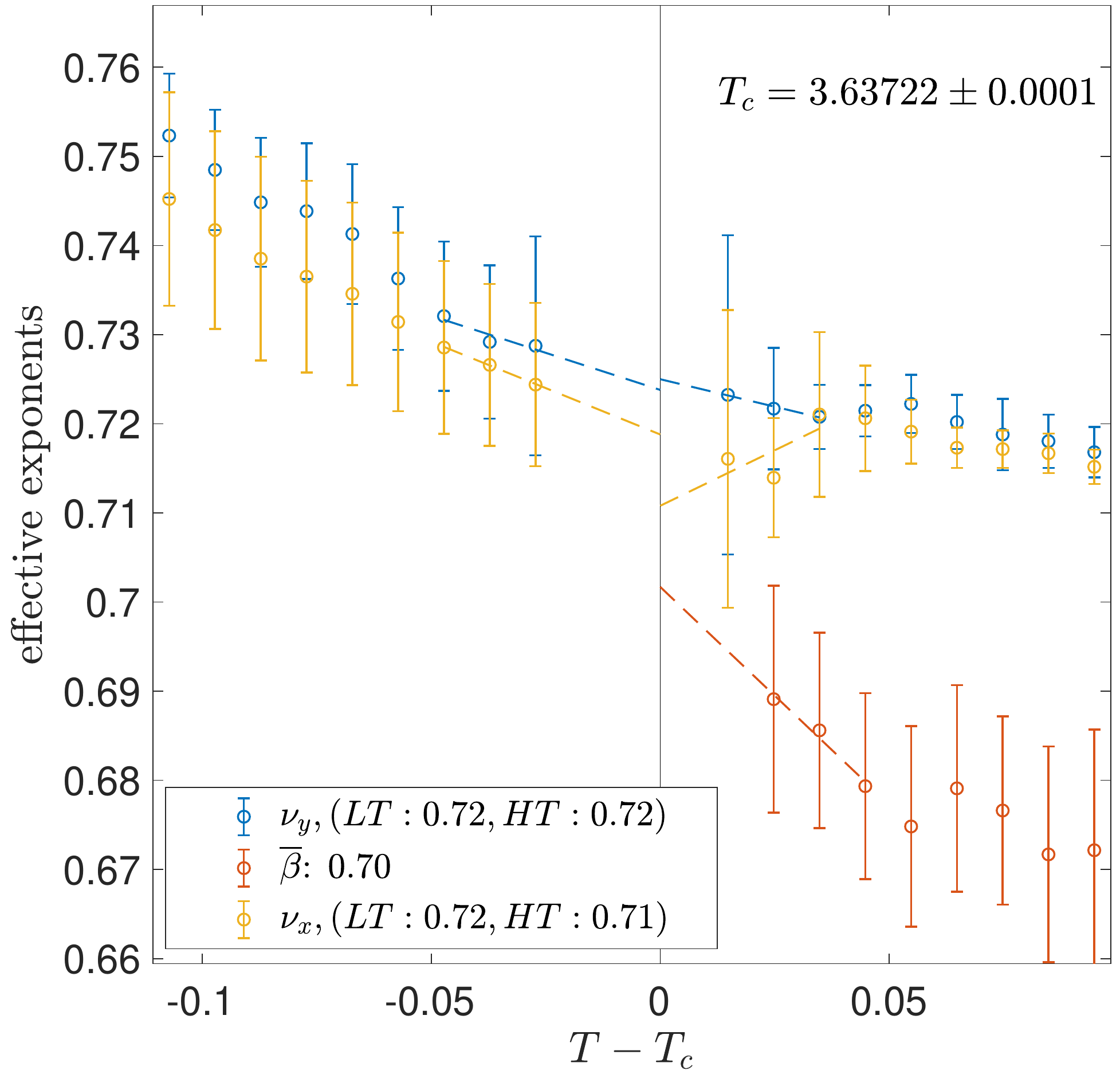}
\caption{Effective exponents as a function of temperature for $\Delta = 0.1$ at $\lambda = 1$. Simulations were made with bond dimension between 100 and 200. The main part of the error bars is due to the extrapolations error. Extrapolated exponents come from a linear fit done on the three last points (dashed lined). We obtain, $\nu_x^{HT} = 0.71\pm0.02, \nu_x^{LT} =0.72\pm0.01$ and $\bar{\beta} = 0.70\pm 0.02$. }
\label{fig:Chiral01}
\end{figure}

We now move to $\lambda = 1$ and small values of $\Delta$. We recall that at this point, the chiral perturbation is irrelevant and the transition should initially be a unique one in the AT universality class. Yet, we do not see any incommensurate melting in the AT universality class. This is probably due to the fact that the parameter range of the AT transition is too small to be detected by our approach. 
The results are summarised in Fig.\ref{fig:Chiral03} and \ref{fig:Chiral01}. Based on those, we exclude the possibility of a two-step transition. Indeed, fixing the critical temperature up to $10^{-4}$ with a unique exponent $\nu_y$ fixes a unique exponent $\nu_x$ with good accuracy. We note that both exponents are smaller than one and the transition must then be unique.

For $\Delta = 0.3$, we obtained after extrapolation $\nu_x^{HT}= 0.692\pm 0.012 $ and $\nu_x^{LT} = 0.710\pm  0.010$ in reasonable agreement with each other. Furthermore, we also extrapolated $\bar{\beta} = 0.689\pm  0.010$. Then, we obtain $\nu_x^{HT} = \bar{\beta}$ within a 0.4\% difference, and an anisotrope exponent $z >1$. Those two features are the main characteristics of the chiral universality class and give already strong evidence in favor of a chiral transition. For $\Delta = 0.1$, we extrapolate from the disordered phase $\nu_x = 0.71\pm0.02$ and $\nu_y = 0.72\pm 0.03$, close to the effective exponents obtained at the Potts point. Similar exponents are obtained from the ordered phase as shown in Fig. \ref{fig:Chiral01}. This is probably due to strong crossovers. However, in contrast to the Potts point we extrapolate $\bar{\beta}= 0.70\pm 0.02$. Such drastic change of behaviour allows us to conclude that the nature of the transition must be different. Once again, we note that $\nu_x^{HT} = \bar{\beta}$ within 1.4\% and in agreement with a chiral transition. 

% extrap on 4pts
% nuyHT \in [0.7519, 0.7247] -> [0.725, 0.752] 0.738 \pm  0.014
% nuyLT  \in [0.7504, 0.7156] -> [0.716, 0.750] 0.733 \pm  0.017
% nuxHT \in [0.6795, 0.7038] -> [0.680, 0.704] 0.692 \pm  0.012
% nuxLT  \in [0.7014, 0.7195] -> [0.701, 0.720] 0.710 \pm  0.010
% beta     \in [0.6790, 0.6981] -> [0.679, 0.698] 0.689 \pm  0.010

% extrap on 3pts
% beta = -0.4881*x + 0.7017 \in [0.72, 0.69]
% nuxHt =  0.249 * x + 0.7108 \in [0.73, 0.69]
% nuxLT = -0.2081*x + 0.7188 \in [0.73, 0.71]
% nuyHT = -0.1233*x + 0.725 \in [0.75, 0.70]
% nuyLT = -0.1664*x + 0.7238 \in [0.74, 0.71]

As already mentioned, just beyond the Lifshitz point, we will obtain $\nu_y>1$ if we try to locate a unique transition due to the KT nature of the upper transition. We can therefore locate the Lifshitz point with the condition $\nu_y = 1$, leading to a Lifshitz point at $\Delta_L = 0.705\pm0.01$.

\subsection{Phase Diagram}

\begin{figure}[t!]
\includegraphics[width=0.48\textwidth]{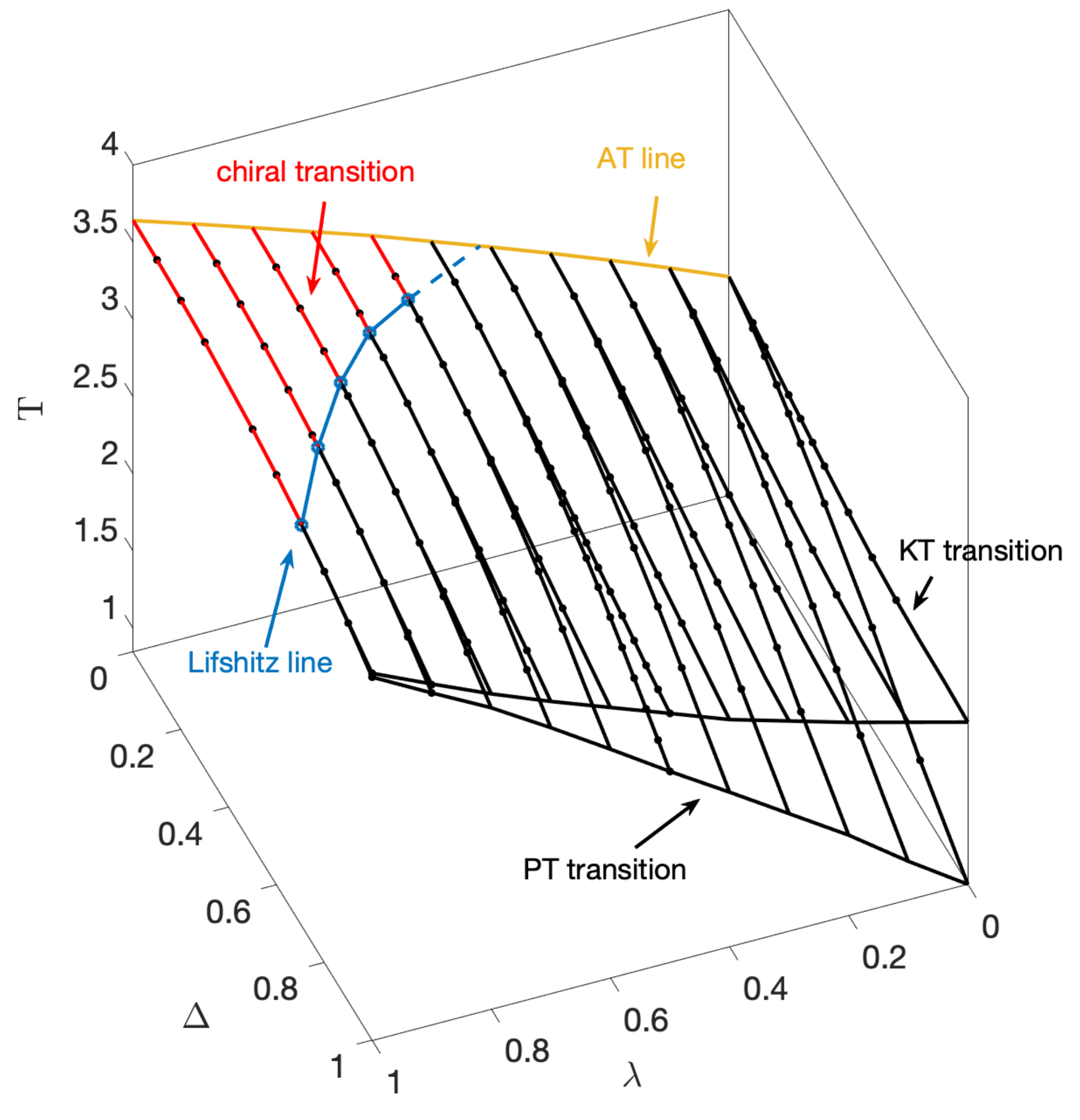}
\caption{Phase diagram of the chiral Ashkin-Teller model. The orange line is the Ashkin-Teller transition in the absence of a chiral perturbation, while the blue line is the Lifshitz transition. The chiral transition is indicated by red lines. Beyond the Lifshitz line, the commensurate-incommensurate transition becomes a two-step process.}
\label{fig:PDCP}
\end{figure}

\begin{figure}[t!]
\includegraphics[width=0.45\textwidth]{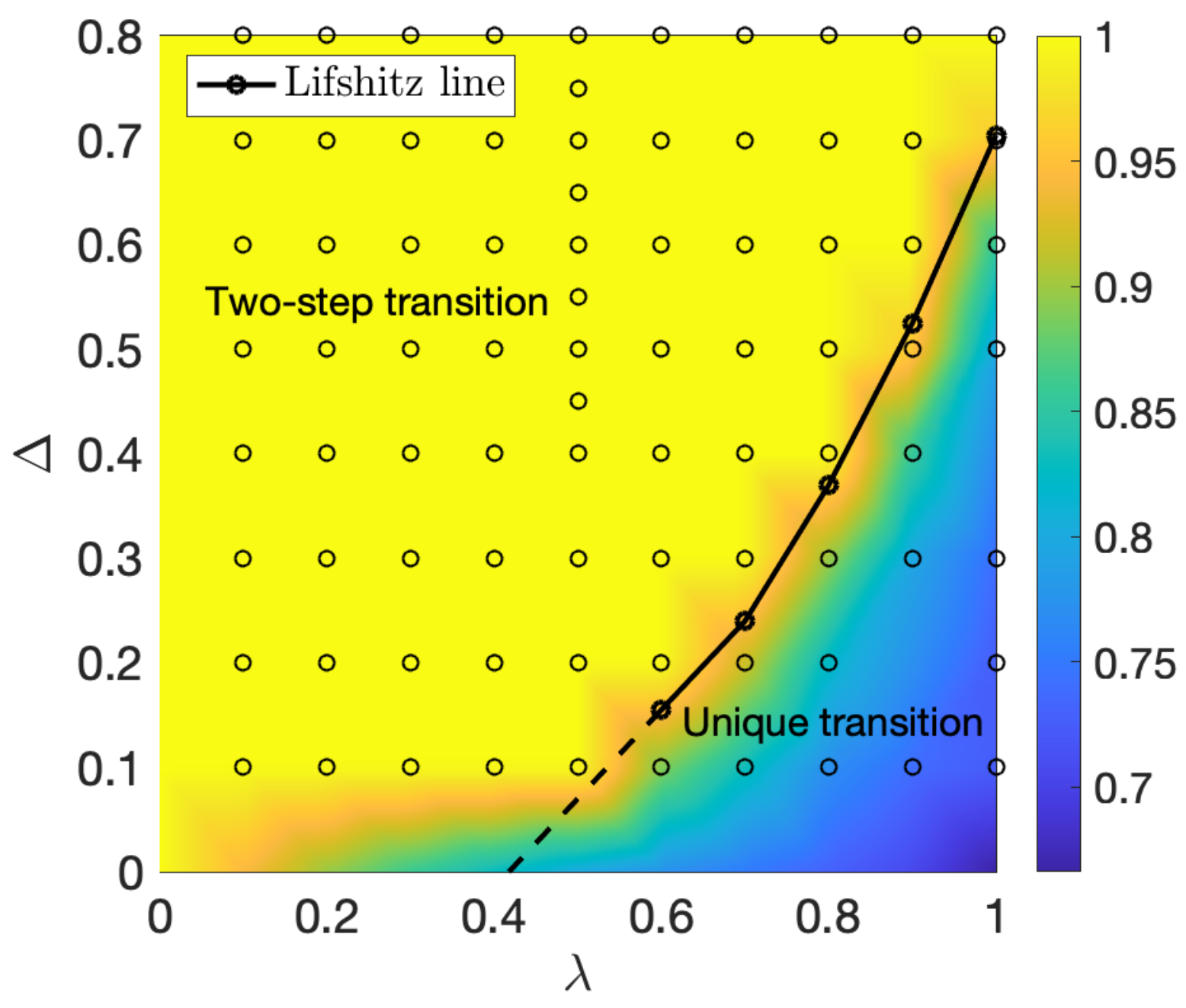}
\caption{Projection of the phase diagram of the chiral Ashkin-Teller model on the $(\Delta,\lambda)$ plane. The color map shows the critical exponent $\nu_y$ from the ordered phase. It is obtained from a linear interpolation of simulations performed at intervals of length 0.1 in the $\Delta$ and $\lambda$ axis (grey dots). In the two-step transition regime, $\nu_y = 1$ by definition due to the PT nature of the transition. The values along the $\Delta = 0$ line are exact. The dotted line is a linear extrapolation of the Lifshitz points at $\lambda = 0.6$ and $\lambda = 0.7$. It extrapolates to $\lambda \simeq 0.42$.}
\label{fig:PDup}
\end{figure}

By applying the same methodology systematically we have been able to map the whole phase diagram, and to identify a line of Lifshitz points. 
This leads to three different regimes depending on the value of $\lambda$. For $0\leq\lambda< \lambda_{c_2}$, the chiral perturbation opens right away a floating phase. By contrast, for $\lambda_{c_2}<\lambda<\lambda_{c_1}$, the chiral perturbation leads to a unique transition in the chiral universality class, and a floating phase only opens for a non-zero value of $\Delta$. Finally, for $\lambda_{c_1}<\lambda\leq 1$, the transition should first be in the AT universality class, a regime too small to be detected numerically. For small but-non zero value of $\Delta$, the transition should become a unique transition in the chiral universality class, and finally, for larger $\Delta$, a floating phase should open. These results are summarised in Fig.\ref{fig:PDCP}, where we show the three dimensional phase diagram, and in Fig.\ref{fig:PDup}, where we show a projection of the phase diagram on the $(\Delta, \lambda)$ plane. 
The similarity between the phase diagram of Fig.\ref{fig:PDCP} and Fig. 12 of Ref.~[\onlinecite{HuseFisher1984}] is striking.
The transition is clearly unique for a range of $\lambda$ below the Potts point $\lambda=1$, and a two-step one for a range of $\lambda$ above the clock-Ising point $\lambda=0$. A linear extrapolation of the Lifshitz point at $\lambda = 0.6$ and $\lambda = 0.7$ towards $\Delta=0$ leads to the estimate
$\lambda_{c2}\simeq 0.42$,  but the curvature of the line of Lifshitz points in Fig. \ref{fig:PDup} suggests that this value should be considered as an upper bound for $\lambda_{c_2}$.  This result for $\lambda_{c_2}$ is in agreement with recent DMRG simulations on the quantum version of the model\cite{Chepiga2021}, for which a chiral transition has been identified in a model with $\lambda \sim 0.57$. %To obtain a lower bound for $\lambda_{c_2}$ would be more tricky and too speculative


%\gtrsim
%\lesssim

\subsection{Specific heat and Kibble-Zurek exponent}

% delta = 0.1, beta = 0.71 , nuy = 0.72
% delta = 0.3 , beta = 0.689, nuy = 0.738
We now discuss and analyse the specific heat exponent $\alpha$ obtained along the line $\lambda = 1$ and its relation with the Kibble-Zurek exponent. We recall the hyper-scaling relation $\nu_x + \nu_y = 2 -\alpha$. If such a relation applies, there are two different ways to determine if, as reported earlier for the three-state Potts model\cite{nyckees}, $\alpha$ is constant along the transition: Either through the direct computation of $\alpha$ itself, or through the computation of the sum $\nu_x + \nu_y$. 

\begin{figure}[t!]
\includegraphics[width=0.45\textwidth]{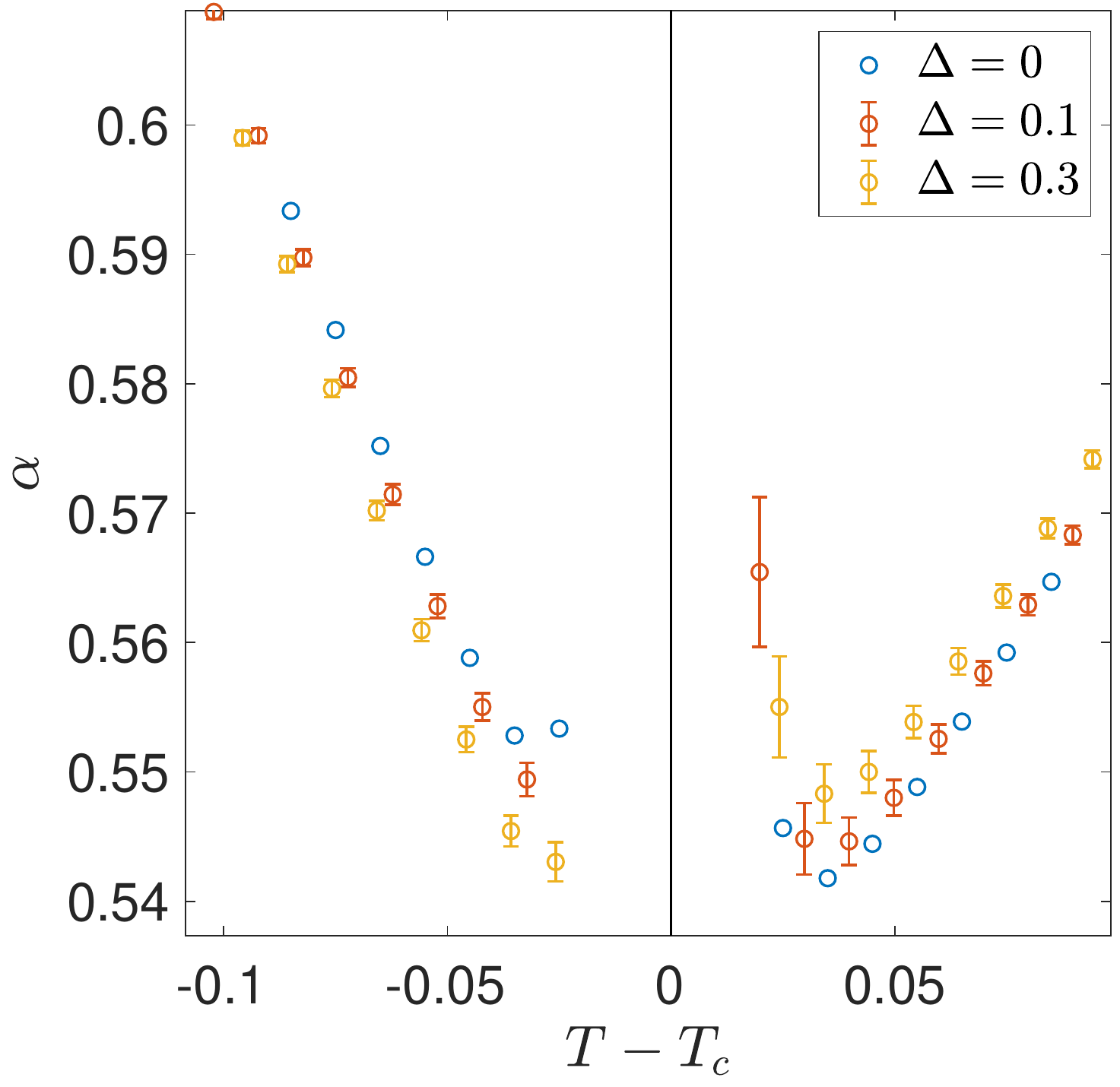}
\caption{Effective specific heat exponent $\alpha_\text{eff}$ for different values of $\Delta$ along the transition at $\lambda = 1$. The simulation are performed at finite $\chi = 200$. The error bars for $\Delta = 0.1$ and $\Delta = 0.3$ come from the uncertainty of the critical temperature. }%The energy has converged for $\Delta = 0$, $\Delta = 0.1$ and $\Delta = 0.3$ to the order of $10^{-9}$, $10^{-8}$ and $10^{-7}$ respectively. }
\label{fig:alpha}
\end{figure}

For the Potts point, the exact value is known to be $\alpha = 2/3$. Due to logarithmic corrections in the correlation length, we have found $2\nu =  1.426$ which gives $\alpha = 0.57$. If we make the hypothesis of a chiral transition, we can use the identity $\bar{\beta} =\nu_x$ and for $\Delta = 0.1$ and $\Delta = 0.3$, we respectively found $\nu_y^{HT} + \bar{\beta} = 1.42$ and 1.427. We also directly studied $\alpha_\text{eff}$ (see Fig. \ref{fig:alpha}) and found evidence of a constant exponent along the transition at $\alpha \simeq 0.55$, which establishes the validity of the hyper-scaling relation. Thus, we observe that $\alpha$ seems to be more or less constant along the chiral transition, as already observed in the $p=3$\cite{nyckees} case and in a  quantum version of the model at $\lambda\simeq 0.57$\cite{Chepiga2021}. We note that a constant value of $\alpha = 2/3$ along the transition and in particular at $\Delta = 0$ would imply that there are still logarithmic corrections in the correlation length along the chiral transition. Thus, we are probably overestimating the sum $\nu_x + \nu_y$, and in fact most certainly $\nu_x$ since, in order to keep $\nu_x + \nu_y  = 4/3$ with $\nu_y>\nu_x$,  $\nu_x$ should be smaller than 2/3. 

The Kibble-Zurek exponent can be measured in arrays of Rydberg atoms\cite{lukin2019}. It is defined as $\mu = \nu/(1+\nu z)$, with $\nu$ the correlation length exponent along the chain and $z$ the dynamical exponent. In our case, $z$ plays the role of the anisotropy exponent $z = \nu_y/\nu_x$, and $\nu$ is the correlation length exponent in the $x$-direction. Therefore, we can compute the Kibble-Zurek exponent as $\mu = \nu_x/(1+\nu_y)$. For reasons discussed above, we choose to express $\mu$ with respect to $\nu_y$. Then, using the hyper-scaling relation and the hypothesis of a constant $\alpha = 2/3$ along the transition, the Kibble-Zurek exponent becomes $\mu = (4/3 - \nu_y) /(1+\nu_y)$. Using this expression of the exponent and the critical exponents obtained from the disordered phase for $\lambda=1$, we found $\mu = 0.36$ and $\mu = 0.34$ for $\Delta = 0.1$ and $\Delta = 0.3$ respectively. These results indicate that the Kibble-Zurek exponent is smaller along the chiral transition than at the Potts point, where $\mu = 0.4$. These results are in qualitative agreement with recent experiments\cite{lukin2019} on chains of Rydberg atoms, which have been argued to be described by the AT model with $\lambda$ close to 1 in the vicinity of the $p=4$ ordered phase\cite{Chepiga2021}, and for which values $\mu \simeq 0.25$ have been reported\cite{lukin2019} . %Thus, we are over estimating the sum $\nu_x + \nu_y$.
% This could be hiding a larger exponent $\nu_y$, then one could obtain a Kibble-Zurek exponent closer to the experimental results.

\section{Results for $\lambda \notin [0,1]$}

%We now comment the properties of the model for $\lambda$ taking different values outside the $[0,1]$ interval.
Since our main purpose was to identify a possible chiral transition we have concentrated in the preceding section on the interval $\lambda\in[0,1]$ since it contains the relevant parameter range $\lambda_{c2}\lesssim 0.32 \leq \lambda < \lambda_{c1}=0.9779$. For completeness we now comment on the effect of a chiral perturbation outside this interval.
\subsection{ $-1 < \lambda < 0$}

In this range, the chiral perturbation stays relevant and the floating phase open up as soon as the chirality is introduced. The zero temperature ground state stays four times degenerate and we do not expect any qualitative change from the $\lambda < \lambda_{c2}$ case.

\subsection{ $\lambda < -1$}

\begin{figure}[t!]
\includegraphics[width=0.48\textwidth]{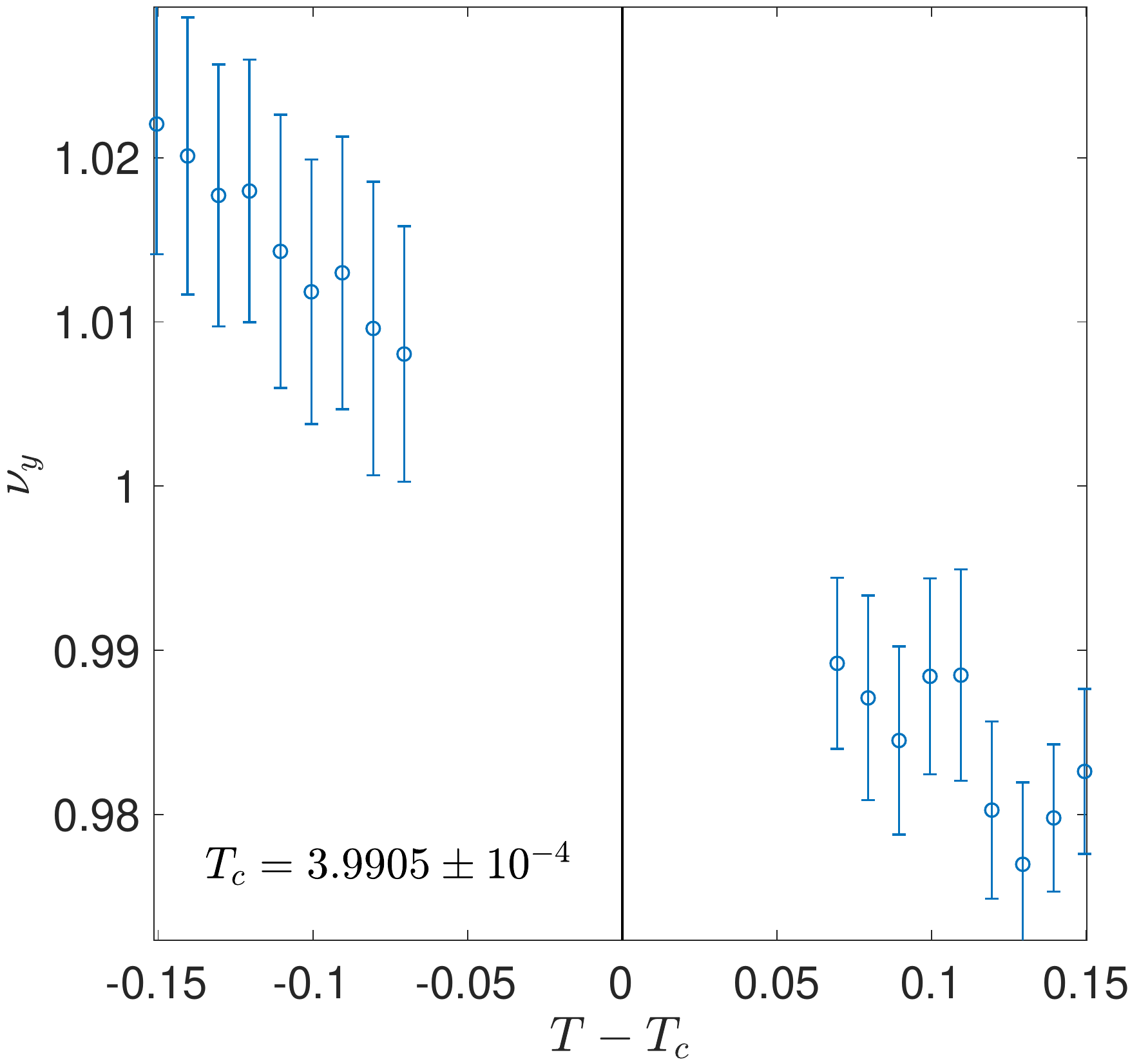}
\caption{Evidence of a unique transition at $\lambda=-2$ and $\Delta=0.1$ with a critical exponent $\nu$ in good agreement with the Ising universality class $\nu = 1$. For this simulation, we have changed the isometries and used those in\cite{orus}, which we noticed to perform better in the anisotropic case.}
\label{fig:L2d1}
\end{figure}

In this case, at $\Delta = 0$, the nature of the zero temperature ground state has changed and so has the transition. It is expected to be described by Ising critical exponents. Indeed, in the $\lambda \rightarrow \infty$ limit, the model is described by Ising variables $u_i = \tau_i \sigma_i$ and the hamiltonian reduces to $H = \sum_{i\sim j} u_i u_j$. The zero temperature ground state is infinitely degenerate and we recover a $\log(2)$ residual entropy. Introducing the chirality will lift the residual entropy but the zero temperature ground state stays  $2^{L}$ degenerate for a $L\times M$ lattice size, and we expect the transition to stay a C-C one. We thus also expect the transition to stay in the Ising universality class. Other values of $\lambda$ within the interval should be described by the same physics. We show results for $\lambda = -2$ at finite chirality $\Delta = 0.1$ in Fig. \ref{fig:L2d1}. And as expected we found a unique $\pi$ commensurate - $\pi$ commensurate transition with a correlation length exponent consistent with the Ising value $\nu= 1$.

\subsection{ $ \lambda > 1$}

\begin{figure}[t!]
\includegraphics[width=0.48\textwidth]{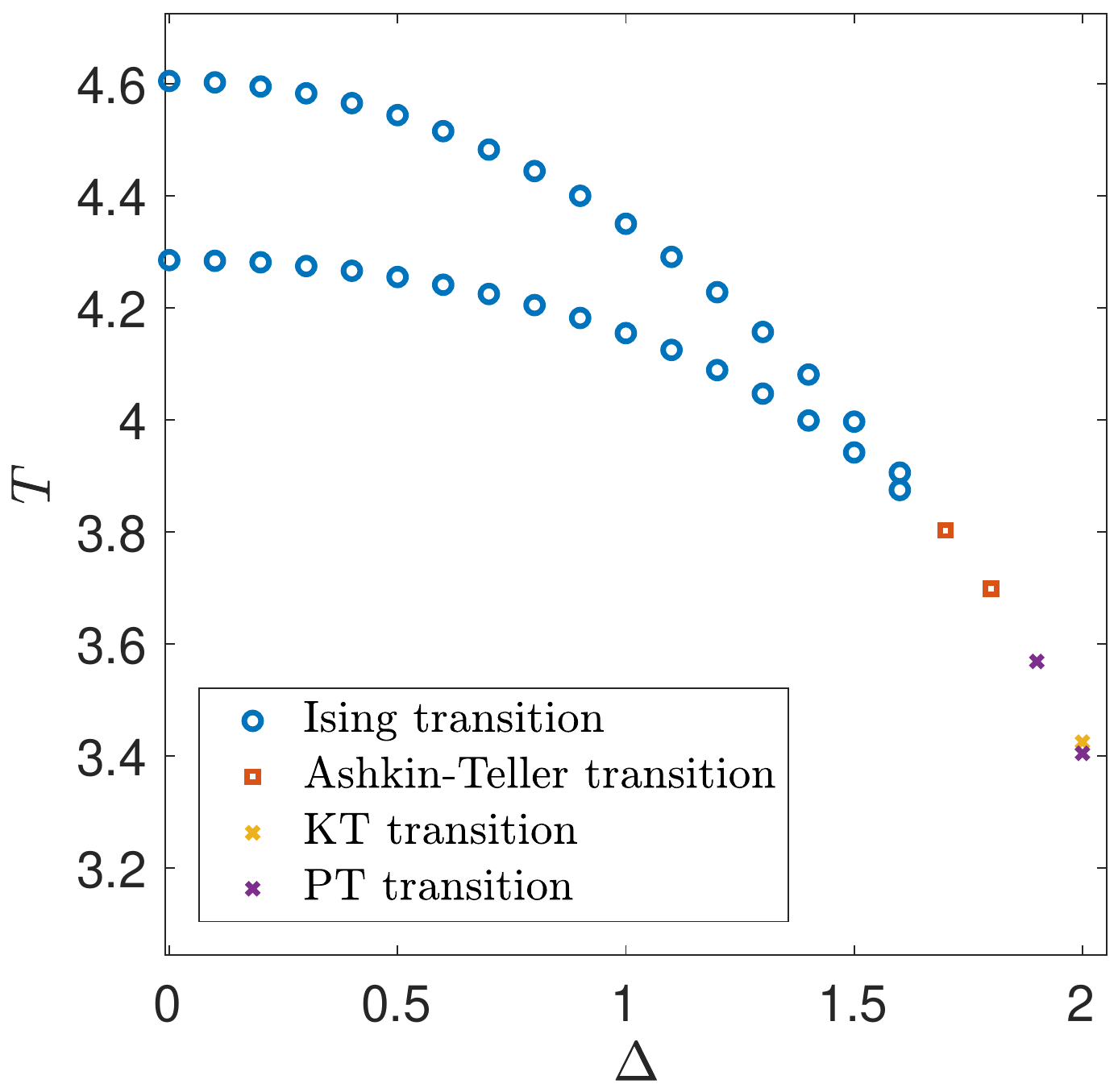}
\caption{Phase diagram at $\lambda=1.7$ (above the Potts point) as a function of $\Delta$. The two Ising transitions at small $\Delta$ merge around $\Delta\simeq 1.7$ into a single phase transition that must be in the AT universality class since the high temperature phase remains commensurate for some range of $\Delta$. At larger $\Delta$, a floating phase bounded by KT and PT transitions develop. For this value of $\lambda$, the region of chiral transition (if any) is too small to be detected.}
%We found two Ising transitions for small $\Delta$. Then a unique transition which we identify to be Ashkin-Teller as it cannot be chiral due to the high temperature being commensurate for $\Delta = 0.18$. At large chirality we recover a two-step transition. In contrast to the $\lambda = 1.05$ case, we cannot say if the chiral transition still exists over a range of parameter between $\Delta \in [0.18,0.19]$.}
\label{fig:PD_Lamb17}
\end{figure}

\begin{figure}[t!]
\includegraphics[width=0.48\textwidth]{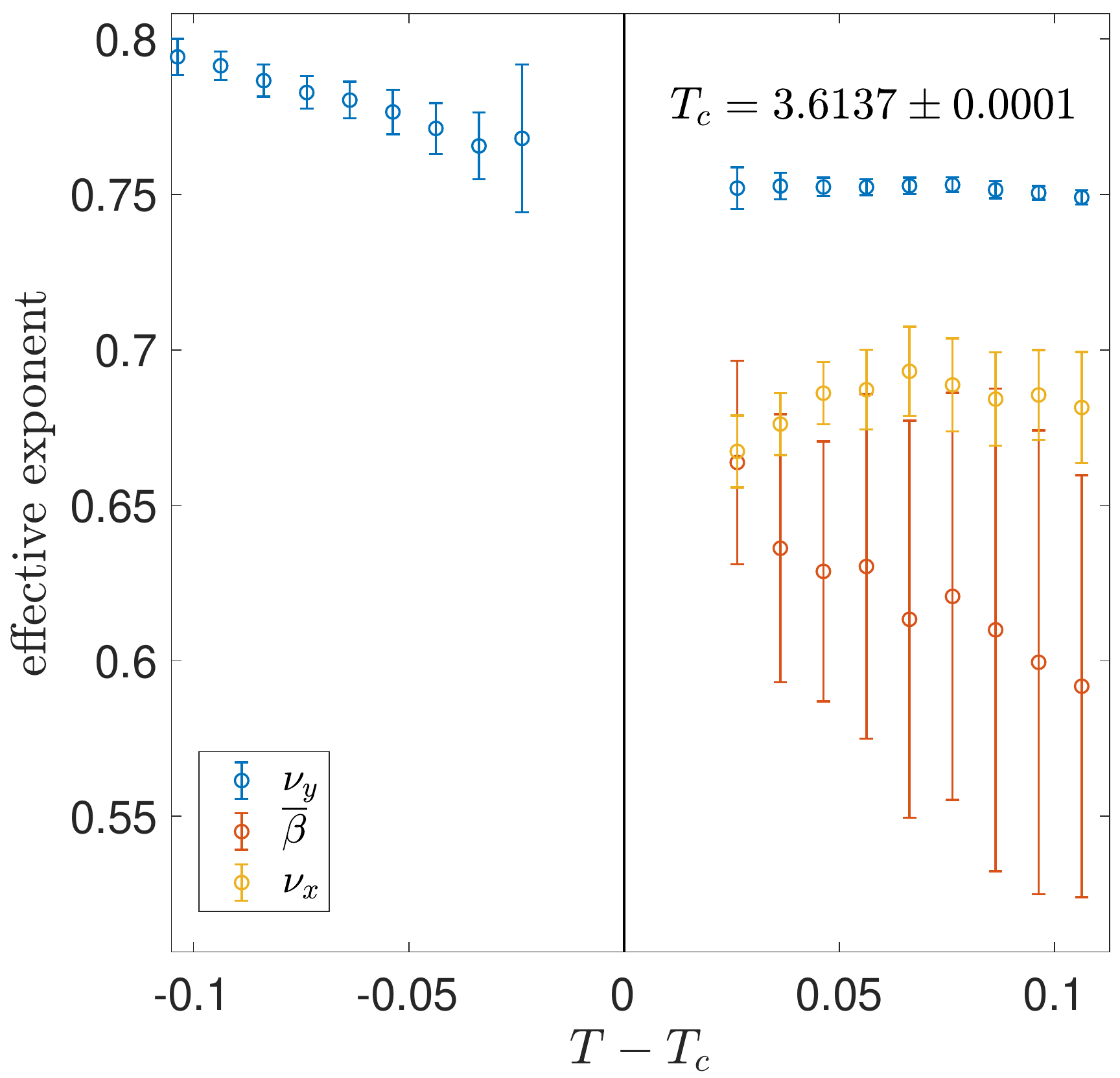}
\caption{Evidence of a unique transition with $\bar{\beta} \simeq \nu_x$ for $\lambda=1.05$ and $\Delta=0.5$, in agreement with a chiral transition.}
\label{fig:L105D05}
\end{figure}

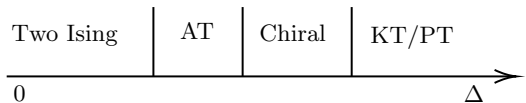
\begin{figure}
\tikzset{every picture/.style={line width=0.75pt}} %set default line width to 0.75pt        
\begin{tikzpicture}[x=0.75pt,y=0.75pt,yscale=-1,xscale=1]
%Straight Lines [id:da42831759358524013] 
\draw    (100,123) -- (355,123) ;
\draw [shift={(357,123)}, rotate = 180] [color={rgb, 255:red, 0; green, 0; blue, 0 }  ][line width=0.75]    (10.93,-3.29) .. controls (6.95,-1.4) and (3.31,-0.3) .. (0,0) .. controls (3.31,0.3) and (6.95,1.4) .. (10.93,3.29)   ;
%Straight Lines [id:da2933762819393073] 
\draw    (174,123) -- (174,88) ;
%Straight Lines [id:da24781074926467883] 
\draw    (219,123) -- (219,88) ;
%Straight Lines [id:da8670816583423626] 
\draw    (274,123) -- (274,88) ;
\draw (329,126.4) node [anchor=north west][inner sep=0.75pt]    {$\Delta $};
\draw (102,126.4) node [anchor=north west][inner sep=0.75pt]    {$0$};
\draw (101,95.4) node [anchor=north west][inner sep=0.75pt]    {$\text{Two Ising}$};
\draw (186,94.4) node [anchor=north west][inner sep=0.75pt]    {$\text{AT}$};
\draw (226,95.4) node [anchor=north west][inner sep=0.75pt]    {$\text{Chiral}$};
\draw (282,95.4) node [anchor=north west][inner sep=0.75pt]    {$\text{KT/PT}$};
\end{tikzpicture}
\caption{Sketch of the expected generic phase diagram for $\lambda>1$. The high-temperature phase above the  Ising and AT transitions is commensurate. 
%The length of the intervals describing the phases are qualitative and are not represented by the right proportion. 
The width of the various phases varies with $\lambda$, and the widths of the sketch are arbitrary. Upon increasing $\lambda$, the width of the chiral transition becomes smaller and smaller to the point where we cannot say for sure if it still exists. }
\label{fig:PDabove1}
\end{figure}

At $\Delta = 0$, for $\lambda>1$, the self dual line does not represent the critical line anymore and the Ashkin-Teller transition splits up into two Ising transitions with in between a nematic phase characterised by $\langle \tau_1 \sigma_1\rangle\neq 0$. When introducing the chirality, the two Ising transitions first stay in the Ising universality class but as the chirality increases, the nematic phase closes and the two Ising transitions merge into a unique one bounded by commensurate phases, as illustrated in Fig.\ref{fig:PD_Lamb17} for $\lambda=1.7$. Upon further increasing the chirality $\Delta$, the high temperature phase becomes incommensurate, and for large enough $\Delta$ a floating phase bounded by KT and PT transitions appears. For this value of $\lambda$, the range of chiral transition (if any) is too small to be detected.  By contrast, for not too large $\lambda$, we found evidence for the existence of a chiral transition to follow up on the AT one. This result is illustrated in Fig. \ref{fig:L105D05}. In the high chirality limit, one recovers a two-step transition separated by a floating phase (not shown). Based on the results illustrated in Fig. \ref{fig:PD_Lamb17} and \ref{fig:L105D05}, a sketch of the expected generic phase diagram as a function of $\Delta$ is shown in Fig. \ref{fig:PDabove1}. Whether the chiral transition persists for all $\lambda$ is an open question that would require further investigation.%, but we do not found evidence of the latter for large $\lambda$'s.

\section{Summary}

Based on the qualitative study of effective exponents, we have been able to give a self consistent picture of the critical exponents and universality classes, and to identify two different types of commensurate-incommensurate transitions, either a two-step one through a KT and a PT transition, or a unique chiral transition. If we take for granted that for $\lambda_{c_1}\simeq 0.9779\leq \lambda \leq 1$, the transition should remain AT for a while, something we have not been able to show numerically because this range is too small, we are led to the conclusion that the Ashkin-Teller family of models can be classified into three regimes according to the way they react to a chiral perturbation: (i) A floating phase opens right away for $0\leq \lambda <\lambda_{c_2}$; (ii) The transition is unique in the chiral universality class for $\lambda_{c_2}\leq \lambda <\lambda_{c_1}$; (iii) The transition remains in the AT universality class before becoming chiral for $\lambda_{c_1}\leq\lambda\leq 1$. The critical value for $\lambda_{c_2}$ could not be pinned down very precisely, but it can be expected to satisfy $\lambda_{c_2}\lesssim 0.42$. We further found evidence for the chiral transition to be characterised by a specific heat exponent that keeps its AT value. In many models, the melting of an ordered period-4 phase into incommensurate phase takes place and the nature of the transition is still debated. If an Ashkin-Teller point is present along the transition, by identifying its universality class and its exponent $\nu$, our prediction can be used to determine the nature of the melting close to that point. Further study on such a model and in particular on the quantum chiral Ashkin-Teller model would be of particular interest to test these predictions.

%Further study on such a model and in particular on the quantum chiral Ashkin-Teller model would be of particular interest to test these predictions. 

{\it Acknowledgments.} We thank Natalia Chepiga and Jeanne Colbois for useful discussions.
This work has been supported by the Swiss National Science Foundation.
The calculations have been performed using the facilities of the Scientific IT and Application Support Center of EPFL.

%merlin.mbs apsrev4-1.bst 2010-07-25 4.21a (PWD, AO, DPC) hacked
%Control: key (0)
%Control: author (72) initials jnrlst
%Control: editor formatted (1) identically to author
%Control: production of article title (-1) disabled
%Control: page (0) single
%Control: year (1) truncated
%Control: production of eprint (0) enabled

%\bibliographystyle{apsrev4-2}
%\bibliography{bibliography,comments}

%apsrev4-2.bst 2019-01-14 (MD) hand-edited version of apsrev4-1.bst
%Control: key (0)
%Control: author (8) initials jnrlst
%Control: editor formatted (1) identically to author
%Control: production of article title (0) allowed
%Control: page (0) single
%Control: year (1) truncated
%Control: production of eprint (0) enabled
%

%merlin.mbs apsrev4-1.bst 2010-07-25 4.21a (PWD, AO, DPC) hacked
%Control: key (0)
%Control: author (72) initials jnrlst
%Control: editor formatted (1) identically to author
%Control: production of article title (-1) disabled
%Control: page (0) single
%Control: year (1) truncated
%Control: production of eprint (0) enabled

\end{document}